\documentclass{aa501}
\usepackage{times}
\usepackage{graphics}
\usepackage{xspace}
\usepackage{epsfig}
\usepackage{jwaabib}
\usepackage{rotating}
\usepackage{dcolumn}

\newcommand{\vsini}{\ensuremath{v \sin{i}}\xspace}

\newcommand{\lgLx}{\ensuremath{\lg{L_{\rm x}}}\xspace}
\newcommand{\lgvsini}{\ensuremath{\lg{(v \sin{i})}}\xspace}

\newcommand{\lgLrat}{\ensuremath{\lg{(L_{\rm x}/L_{\rm bol})}}\xspace}
\newcommand{\lgTeff}{\ensuremath{\lg{T_{\rm eff}}}\xspace}

\begin{document}


\title{X-ray emission from young stars in Taurus-Auriga-Perseus: \\
Luminosity functions and the rotation - activity - age - relation\thanks{Tables~2~to~7 are available only electronically.}}

\author{B. Stelzer\inst {1} \and R. Neuh\"auser\inst {1}} 

\institute{Max-Planck-Institut f\"ur extraterrestrische Physik,
  Postfach 1312, 
  D-85741 Garching,
  Germany} 

\offprints{B. Stelzer}
\mail{B. Stelzer, stelzer@xray.mpe.mpg.de}
\titlerunning{X-ray emission in Taurus-Auriga-Perseus}

\date{Received $<$date$>$ / Accepted $<$date$>$}

\abstract{
We report on a systematic search for X-ray emission from pre-main sequence
and young main sequence stars in the Taurus-Auriga-Perseus region.
Our stellar sample consists of all T Tauri stars from the
Taurus-Auriga region, and all late-type stars from the Pleiades and
Hyades clusters which have been observed by the {\em ROSAT} PSPC in pointed
observations.
We present the X-ray parameters for all observed stars
in tables. Next to the basic results of the data analysis 
(such as count rates,
exposure time, and off-axis angle) we give X-ray luminosities and
hardness ratios for all detected stars. Upper limits are given for 
non-detections. Detection rates for different spectral types are compiled.
We use these results to 
study the connection between coronal X-ray activity and stellar parameters
for different subgroups of our sample. In particular we 
compile X-ray luminosity functions (XLF), 
and discuss the relations between X-ray emission and 
spectral type, age, and rotation, which have been disputed extensively in
the past. Here, we study these questions with the largest sample so far.
The XLF for classical and weak-line T Tauri stars are different, with
weak-lines being the stronger X-ray emitters.  Proceeding towards the
main-sequence (Pleiades, Hyades) the X-ray luminosity declines for all
spectral types examined (G, K, and M stars).
Within an age group $L_{\rm x}$ decreases towards later spectral types,
while $L_{\rm x}/L_{\rm bol}$ remains constant or even increases,
reflecting the opposed influence of stellar radius, i.e. emitting area, and
convection zone depth. For 
a given spectral type the fastest rotators show the highest X-ray luminosity.
Rotation rate and X-ray emission are clearly correlated for all groups of
stars with power law indices for \lgLrat versus $\lg{P_{\rm rot}}$ of
$\sim -0.7$ to $-1.5$. The study of XLF for binary stars shows that
the known unresolved secondaries likely contribute a significant amount 
to the X-ray emission.
}

\maketitle

\section{Introduction}\label{sect:intro}

Late-type stars are known to sustain a dynamo which is powered by the
combination of convective motions and rotation. The resulting magnetic
field is thought to be responsible for various observational 
phenomena commonly referred to as `activity'. Stellar activity indicators
are pervasive in all layers of the atmospheres of late-type stars. The 
best-studied magnetic field tracers include chromospheric H$\alpha$ and 
Ca II emission lines and coronal X-rays. 

Strong and variable X-ray emission 
is observed from the early pre-main sequence (PMS)
T Tauri stage to main sequence flare stars. Comparative studies of
the emission levels of young stars at different ages 
may shed light on the origin and evolution of magnetic activity, which may
be linked to angular momentum evolution.

T Tauri Stars (TTS) are divided into two subclasses according to the strength
of their H$\alpha$ emission. Classical TTS (cTTS) are characterized by
strong H$\alpha$ emission, while in weak-line TTS (wTTS) the equivalent
width of H$\alpha$ is smaller than $10\,{\rm \AA}$. 
In contrast to cTTS, wTTS do not show obvious 
signs of accretion and optically thick disks, 
and therefore are thought to represent a later
evolutionary stage at which the circumstellar disk has become optically
thin or dispersed. Nevertheless, some wTTS occupy the same region in 
the Hertzsprung-Russell diagram as the cTTS (see e.g. \cite{Walter88.1}, 
\cite{Alcala97.1}), i.e. there seems to be no preferred disk lifetime between
several $10^5$ and $10^7\,{\rm yrs}$.

During the PMS contraction stars gain angular momentum and should spin up.
However, cTTS show much slower rotation rates than wTTS
(\cite{Bouvier93.1}) despite their high accretion rates. This can be explained
by magnetic coupling between the star and the disk in cTTS which allows
to regulate angular momentum transport without spinning up the star
(\cite{Koenigl91.1}, \cite{Edwards93.1}, \cite{Bouvier97.1}).

The strength of activity
is thought to decline with increasing stellar age. 
\citey{Skumanich72.1} have proposed a $1/\sqrt{t}$-law for  
the decay of stellar activity in young stellar clusters based on Ca\,II
line observations. 
The power-law relation was confirmed by \citey{Feigelson89.1} for a sample 
including PMS objects in Chamaeleon. 
However, the study of \citey{Walter88.1} seemed to indicate that during
the PMS phase the X-ray emission remains nearly constant, while for
ages $\geq 10^8\,{\rm yrs}$ it decays exponentially. 
\citey{Walter91.1} have argued that a power-law decay may be 
an artefact that occurs 
when the X-ray luminosity is used as activity indicator in a sample
composed of stars with different stellar radii.
Using the surface flux \citey{Walter91.1} have shown 
that the decay of various activity diagnostics probing the lower 
chromosphere, transition region, and corona can be described by an 
exponential. 

An evolutionary decay of the X-ray emission is favored also by 
studies of the {\em Einstein} observatory ({\em EO}) 
which have shown that the XLF of stellar clusters 
with different ages are displaced from each other (see
e.g. \cite{Feigelson89.1}, \cite{Damiani95.1}).
However, rather than being a pure age effect, the decreasing activity 
(i.e. the decay of the dynamo efficiency) might be explained
by the slowing down of the rotation with increasing age on the main
sequence (MS).
A connection between the dynamo efficiency and the rotation rate 
is also supported by correlations between the X-ray activity and 
the rotational velocity (see \cite{Pallavicini81.1}, \cite{Bouvier90.1}, 
\cite{Neuhaeuser95.1} = N95)
found to hold for all kinds of active stars, featuring such different
objects like dMe stars, RS~CVn binaries, and TTS.
For the fastest rotators among the Pleiads (\cite{Stauffer94.1}) and among the
wTTS (\cite{Damiani95.2}) the relation flattens out suggesting `saturation'
of the X-ray emission at very high rotation speeds. \citey{Stauffer94.1}
have proposed that the large spread observed in the X-ray luminosities of
slower rotators are
caused when more and more stars leave the saturation level thus increasing 
the dispersion. 
A later study by \citey{Micela96.1} showed no significant correlation
between $L_{\rm x}$ and \vsini for Pleiades stars, indicating that 
rotation can not be the only parameter governing the X-ray emission level of
young stars.
To date, no solution has been found to the question
whether age or rotation determines the level of stellar activity.

If rotation does determine the activity, then cTTS should show lower X-ray
luminosities than wTTS.
X-ray luminosity distribution functions (XLF) provide a statistical
tool to compare the X-ray properties of different stellar samples. But so
far, studies of the XLF of cTTS and wTTS have not led to a conclusive result 
concerning the evolution of X-ray activity during the PMS. 
N95 have shown that X-ray emission increases with age from cTTS to wTTS, but
decreases after the wTTS phase. While N95
found that wTTS in Taurus-Auriga emit significantly more X-rays as compared
to cTTS from the same region, \citey{Feigelson93.1} found little
evidence for systematic differences in the 
XLF of wTTS and cTTS in Chamaeleon, and argued that the 
differences in the X-ray luminosity between wTTS and cTTS in Taurus-Auriga
might be attributed to as yet undicovered wTTS at the low-luminosity
end of the distribution.

In this paper we examine the connection between X-ray activity, age, and
rotation comparing different samples of young stars.
The Taurus-Auriga-Perseus region is well suited for a study of the
evolution of stellar activity 
because it hosts a large number of young stars at different
ages: The molecular clouds of Taurus-Auriga are sites of
ongoing star formation and have
produced many TTS with ages between 
several $10^5$ to $10^7\,{\rm yrs}$. Besides this star forming region,
two young star clusters are located in the area, the Pleiades and
the Hyades, at ages of $10^8\,{\rm yrs}$ and $6\,10^8\,{\rm yrs}$ respectively.

Our sample is larger and the sensitivity improved
compared to all previous studies of this issue. 
The paper presented here is connected to an earlier analysis of 
archived {\em ROSAT} observations (\cite{Stelzer00.1}). 
Here we present a list of X-ray parameters for all known TTS in Taurus-Auriga,
the Pleiades, and the Hyades as observed by {\em ROSAT} during pointed PSPC
observations. Both detected and undetected sources are considered, i.e.
for non-detections upper limits are given.
In Sect.~\ref{sect:obs} we describe the data base 
and data reduction and give the results from source detection.  
In Sect.~\ref{sect:hrds} the Hertzsprung-Russell diagrams (H-R
diagrams) for cTTS and wTTS in Taurus-Auriga are presented.
Our special interest (Sect.~\ref{sect:ldfs}) 
is to compare the XLF of the different
stellar groups with respect to the following issues: (i) Are the
luminosity functions of cTTS and wTTS different, (ii) how does
the X-ray luminosity evolve with stellar age, (iii) how does
it depend on the spectral types of the stars and their binary character.
Furthermore, we will explore the relation between X-ray emission and
rotation rate for the largest sample studied so far 
(Sect.~\ref{sect:x_rot}).
In Sect.~\ref{sect:discussion} we discuss and summarize our results.

\section{Observations}\label{sect:obs}

\subsection{Data Base}\label{subsect:database}

The sky region examined is confined to the Taurus-Auriga-Perseus 
area. A detailed description of this region is given in \citey{Stelzer00.1}
(hereafter SNH00) where we have also presented a sky map showing the 
{\em ROSAT} PSPC observations subject to this and the previous study.
The stellar sample investigated in this paper is identical to the one
described in SNH00. However, we omit stars from the Perseus clouds IC\,348
(\cite{Preibisch96.1})
and NGC\,1333 (\cite{Preibisch97.1}), 
since due to their larger distance the PSPC images are
dominated by source confusion.
We analyse the X-ray emission of young, late-type stars, 
represented by TTS and members from the Pleiades and
Hyades clusters. The selection of the Pleiades and Hyades as examples 
of young clusters was motivated by their spatial vicinity  
to the Taurus-Auriga molecular clouds when projected to the sky. 
For this reason many Pleiads and Hyads lie in the same {\em ROSAT}
PSPC fields. Most of the X-ray detected Pleiads and Hyads are zero-age
main-sequence (ZAMS)
stars. There are also some (higher-mass) post-MS stars which are not
studied here, and many PMS brown dwarfs. 
The sample examined extends down to the latest M-type objects and 
includes brown dwarf candidates. 
Most of these are below the
detection limit. However, we have detected an M9-type
object in Taurus-Auriga, the latest type PMS dwarf seen to emit X-rays so
far (see Sect.~\ref{subsect:m9dwarf}).
The coolest object detected in the Pleiades 
has spectral type M5. In the Hyades we detect objects down to M9 (spectral
types determined from measurements of $B-V$).
With their different ages the three groups of stars (TTS, Pleiads, and
Hyads) allow to examine the evolution of the X-ray luminosity.

We have selected all pointed PSPC observations from the {\em ROSAT}
Public Data Archive available in October 1998 which contain any TTS in the
Taurus-Auriga region, any Pleiad, or any Hyad in the field of view.
The TTS in that area of the sky are part of the Taurus-Auriga
molecular clouds located at $140\,{\rm pc}$ (\cite{Elias78.1}, 
\cite{Wichmann98.1}).
For the distance to the Pleiades cluster we have adopted $116\,{\rm pc}$,
the value given by \citey{Mermilliod97.1}. Those Hyades stars for which no
individual Hipparcos parallaxes are available are assumed to 
be at a distance of $46\,{\rm pc}$ (\cite{Perryman98.1}).

A detailed description of the membership lists for TTS, Pleiads, and
Hyads is given in SNH00. 
SNH00 also have presented a complete list of the pointed 
{\em ROSAT} PSPC observations analysed here. 
In the earlier paper we were dealing with the same observations but have 
concentrated on large X-ray flares observed on 
detected stars. Now we discuss the X-ray characteristics of the whole
sample, including non-flaring stars and non-detections. 
Therefore, we also analyse 
the short exposures and observations with unstable background
marked with an asterisk in Table 1 of SNH00, 
and not considered in that earlier investigation.

\subsection{Source Detection}\label{subsect:soudet}

Source detection is performed based on a maximum 
likelihood method which combines local and map source detection algorithm
(see Cruddace et al. 1988).
Sources with a $ML \geq 7.4$ (corresponding to $\sim 3.5$ Gaussian $\sigma$
and shown to be the best choice by N95) 
are considered to be a detection. The probability for existence of a 
source of given $ML$ is given by $P = 1 - \exp{(-ML)}$.
For $ML = 7.4$ the probability is 0.9994, and among the $\sim 800$ detected 
young stars we would expect to find less than one spurious source.

Observations whose center positions are less than $15^{\prime\prime}$
apart have been merged to increase the sensitivity for faint detections.
The photon extraction radius of the X-ray sources is not well
defined if the off-axis positions in individual observations that are
merged differ strongly from each other. 
Therefore, we have analysed observations
with less overlap, i.e. more than $15^{\prime\prime}$ separation, separately.
The center of the merged image is the center from all individual
observations that are added up. The off-axis positions of X-ray sources
in merged pointings are computed with respect to this averaged pointing
position.

As the positional accuracy of the {\em ROSAT} PSPC declines towards
the edge of the detector, the identification radius between 
optical and X-ray position depends on the off-axis angle of the source.
We have computed the normalized cumulative number of 
identifications as a function of the offset between optical and X-ray 
position, $\Delta_{\rm ox}$, for different ranges of off-axis angles. 
Following N95, for each of these distributions 
we have determined the turnover point, $\Delta_{\rm ox,max}$,
which corresponds to the value of $\Delta_{\rm ox}$ where wrong
identifications begin to contribute significantly to the detected sources.
We have then performed a linear fit
to the mean off-axis angle as a function of this critical offset
$\Delta_{\rm ox,max}$. 
The fit values of $\Delta_{\rm ox,max}$ for all examined
off-axis ranges are listed in Table~\ref{tab:idrad}. These values are used
as identification radii for the cross-correlation of membership lists
and X-ray observations.
For off-axis angles below $27.5^\prime$ a maximum offset between optical 
and X-ray position of $40^{\prime\prime}$ 
is appropriate. Note, that this value agrees with the value found by 
N95 for the {\em ROSAT} All-Sky Survey (RASS).
Sources which are located further than $50^\prime$ from the detector center
are ignored in the analysis presented here, 
because at large off-axis angles the point spread function 
deviates from a Gaussian and can not be adequately modeled by the available
software.
\begin{table}
\caption{Maximum offset allowed between optical and X-ray position $\Delta_{\rm
ox,max}$ for different off-axis angles $\Omega$. $\Delta_{\rm ox,max}$ is the
best-fit value of a linear distribution of offsets found from normalized
cumulative numbers of identifications (see text).}
\label{tab:idrad}
\begin{tabular}{rccclr} 
\multicolumn{5}{c}{Off-axis angle} & $\Delta_{\rm ox,max}$ \\ 
\multicolumn{5}{c}{[arcmin]}       & [arcsec] \\ \hline
       &     & $\Omega$ & $\leq$ & $27.5$ & $40.0$ \\
$27.5$ & $<$ & $\Omega$ & $\leq$ & $30$   & $42.4$ \\
$30$   & $<$ & $\Omega$ & $\leq$ & $32.5$ & $49.5$ \\
$32.5$ & $<$ & $\Omega$ & $\leq$ & $35$   & $56.7$ \\
$35$   & $<$ & $\Omega$ & $\leq$ & $37.5$ & $63.8$ \\
$37.5$ & $<$ & $\Omega$ & $\leq$ & $40$   & $70.9$ \\
$40$   & $<$ & $\Omega$ & $\leq$ & $42.5$ & $78.1$ \\
$42.5$ & $<$ & $\Omega$ & $\leq$ & $45$   & $85.2$ \\
$45$   & $<$ & $\Omega$ & $\leq$ & $47.5$ & $92.3$ \\
$47.5$ & $<$ & $\Omega$ & $\leq$ & $50$   & $99.5$ \\ \hline
\end{tabular}
\end{table}

We have computed the count rates of detected and undetected sources by
integrating all events within a circular region around the source position,
i.e. the X-ray position for detections and the optical position for 
non-detections. We use the 99\% quantile of the point spread function at 
1\,keV as photon extraction radius, except for those few cases where the broad band 
X-ray image shows that the source obviously exceeds this radius 
probably due to the energy being different
from 1\,keV. For these special cases we determine the 
optimum radius individually by visual inspection of the X-ray image. 

The measured counts are background subtracted and divided
by the exposure time obtained from the exposure map to determine 
the count rates. For the background subtraction we have used the
information from the background maps. This method is useful in crowded
fields where a background annulus around the source may easily be
contaminated by adjacent sources.

In the crowded Pleiades region 
occasionally two or more X-ray sources show significant overlap. 
In order to separate the contributions from each star 
we were forced to decrease the photon extraction radius of these
sources. This leads to an underestimation of the true count rate,
but should not effect our results due to the low number of confused
stars ($15$ versus $>\,200$ detections among the Pleiades).

\subsection{Results of Source Detection}\label{subsect:res_soudet}

The result of source detection and identification is summarized in 
six tables: 
Tables~2,3,~and~4 
contain the X-ray parameters of all detected TTS, Pleiads, and Hyads, and 
in Tables~5,6,~and~7
the X-ray characteristics of undetected TTS, Pleiads, and Hyads are listed.

In Tables~2~-~7 the first column contains a number for the observation 
referring to the numbering in 
Table~1 in SNH00. (See SNH00 for the {\em ROSAT} 
observation request numbers.)
For merged observations we give the numbers
of all pointings that have been added up. Column~2 is the designation
of the stars. Column~3 contains two flags, one that gives the type of TTS 
(`W' - wTTS, `C' - cTTS) and another one
for the multiplicity of the stellar system (`S' - single, `B' - binary, `T'
- triple, and `Q' - quadruple). 
The distinction between cTTS and wTTS is based mainly on the standard
H$\alpha$ equivalent width boundary of 10\,\AA~~ together with the
spectral type of the star (i.e. the H$\alpha$ flux), which is similar
to the suggestion by \citey{Martin97.1} to use different $W_{\rm H\alpha}$
boundaries for different spectral types (GKM). Furthermore, we make use of 
indications for circumstellar material as revealed from IR and mm 
observations.
SU\,Aur, e.g., is of spectral type G2 and $W_{\rm H\alpha}$ is between 3.5
and 5\AA~, but it also has a massive disk and, therefore, clearly is a cTTS.
The H$\alpha$ equivalent widths are taken from N95,
\citey{Kenyon95.1}, and \citey{Wichmann96.1}.
The spectral types are shown in 
column~4. The spectral types of Pleiades and Hyades stars
were derived from the $B-V$ measurements given in the Open Cluster Data Base
compiled by C. Prosser and colleagues (and available at
ftp://cfa-ftp.harvard.edu/pub/stauffer/clusters)
using the conversion of \citey{Schmidt-Kaler82.1}.
For TTS in Taurus-Auriga we have adopted the spectral types 
compiled by N95 and \citey{Koenig01.1}.

For all detected stars (Tables~2~-~4) we 
list the X-ray position (columns~5~and~6), 
the offset $\Delta$ between optical and X-ray position (column~7),
the off-axis angle (column~8), and the maximum likelihood (column~9)
of existence.  
We give the X-ray hardness ratios $HR1$ and $HR2$ in columns~10
and~11.
The PSPC hardness ratio $HR1$ is defined as follows:
\begin{equation}
HR1 = \frac{H - S}{H + S}
\end{equation}
where $H$ is the hard band count rate between 0.5-2.0\,keV, and $S$ is
the count rate in the soft band (between 0.1-0.4\,keV). $HR2$ is given
by:
\begin{equation}
HR2 = \frac{H2 - H1}{H2 + H1}
\end{equation}
where $H2$ and $H1$ are the count rates in the upper and lower part of
the hard band between 0.5-0.9\,keV ($H1$) and 0.9-2.0\,keV ($H2$), 
respectively.
In cases where no counts are observed in any one energy band,
and therefore $HR1$ or $HR2$ are either $+1.0$ (no soft counts) or $-1.0$ 
(no hard counts) we have computed upper limits to the hardness from the
counts in the background. 
Column~12 gives the exposure time and column~13 the X-ray luminosity.

In order to determine the count-to-energy-conversion-factor $CECF$ 
for the compilation of luminosities we have 
used the hardness criterion given by \citey{Fleming95.1}: 
$CECF = (8.31 + 5.30 \cdot HR1) \cdot 10^{-12}\,{\rm erg\,cm^{-2}\,cts^{-1}}$. 
Since the soft band in $HR1$ is sensitive to $A_{\rm V}$, this way 
we implicitly take account of the extinction. 
It should be noted that $HR1$ `saturates' 
for extinctions $A_{\rm V} > \sim 0.5$. High extinctions 
are however rare in the Taurus region, and do not play a role for the
Pleiades and Hyades. But to ensure that no systematic errors are introduced by
this method of count-to-energy conversion we have compared the resulting
distribution of X-ray luminosities with those directly derived from the
available $A_{\rm V}$ measurements (see Sect.~\ref{subsect:ldf_rass_point}).

The values of the luminosity given in Tables~2~-~4 have been derived
dividing the count rate by the multiplicity of the stellar system. 
This means we assume  
that each of the components in the system contributes the same level
of X-ray emission (see \cite{Koenig01.1} and 
Sect.~\ref{subsect:ldf_sing_bin}). 
The mean value of the $CECF$ is $1.00 \pm 0.25~10^{-11}\,{\rm
erg\,cm^{-2}\,cts^{-1}}$. 
This value was used to
obtain the luminosity in cases where $HR1$ 
is a upper/lower limit, and therefore
Fleming's relation can not be applied.
Uncertainties in $\lg{L_{\rm x}}$ are derived from the statistical 
errors without taking account of systematic uncertainties in the 
distance estimate. 

X-ray parameters for non-detections are summarized in 
Tables~5~-~7.
The meaning of columns~1~to~4 in Tables~5~-~7
is the same as in Tables~2~-~4.
In columns~5~and~6 we list the optical position.
The off-axis angle of the undetected stars is given in column~7.
Column~8 contains the upper limits to the source counts, 
column~9 the exposure time, 
and column~10 the X-ray luminosity. We have used the mean value of
the $CECF$ for the compilation of an upper limit to $L_{\rm x}$ 
in the case of non-detections.
The X-ray luminosity was divided by the number of stellar components.

Multiple stellar systems are represented by a single entry in 
Tables~2~-~7, but the designations and if 
known the spectral types of all components are given. Whenever more than one 
star lies in the X-ray-to-optical identification radius we list the
designations of all possible counterparts.

If a star was detected in both unmerged and merged observations
we list only the result from the merged observations. The same applies
to stars which are detected neither in the merged nor in the unmerged
observations. Here, we list only the upper limit from the merged
observation. In a few cases a 
star was detected in a single but not in the merged
observations. This can occur if the source is not within the
inner $50^\prime$ of the merged observation due to the shift in 
pointing centers during the merging process, or if the source or 
background is variable.

Stars which have shown an X-ray flare (discussed by SNH00)
are represented by their quiescent emission, i.e. the flare has been 
removed from the data. We have marked the flare observations in 
Tables~2~-~7 by a label after the
observation ID. 

\scriptsize


\addtocounter{table}{+6}

\normalsize

\subsection{Detection Rates}\label{subsect:sptypes}

The aim of this study is to examine the X-ray emission from magnetically
active stars. 
Stars of spectral types earlier than $\sim$ F5 are not
expected to show dynamo activity because they have no or only shallow
convection zones (\cite{Walter83.1}).
We are not interested in the X-ray emission of these 
stars because they obey a different emission mechanism.
Therefore, we restrict the following analysis on stars
with spectral types G and later.

An overview over the detection rates for stars from the different stellar 
groups is given in Table~\ref{tab:det}. 
We have split up each sample according to the spectral types of its
members. 
The column labeled `$N_{\rm D}$' gives the sum of all detections, and `$N_{\rm N}$'
is the number of non-detections. The number of observed stars
(column $N_{\rm stars}$) is smaller than $N_{\rm D} + N_{\rm N}$ due to 
multiple observations of a given sky position. The columns labeled 
`$N_{\rm mult.FOV}$' and `$N_{\rm mult.D}$' denote
the total number of stars observed/detected in more than one observation.
Note, that for most multiple stars only the spectral type of the primary is
known, and therefore the stellar system has only one entry in Table~\ref{tab:det}.
\begin{table}
\caption{X-ray statistics of TTS, Pleiads, and Hyads observed and detected in pointed
{\em ROSAT} PSPC observations. 
$N_{\rm D}$ -- Number of detections, 
$N_{\rm N}$ -- Number of non-detections, 
$N_{\rm stars}$ - Number of different stars observed, 
$N_{\rm mult. FOV}$ -- Number of stars in the FOV of more than one
observation, 
$N_{\rm mult. D}$ -- Number of stars detected in more than one observation.}
\label{tab:det}
\begin{tabular}{lrrrrr} \hline
Sp.Type & $N_{\rm D}$ & $N_{\rm N}$ & $N_{\rm stars}$ & $N_{\rm
mult. FOV}$ & $N_{\rm mult. D}$ \\ \hline
\multicolumn{6}{c}{\bf Taurus-Auriga TTS} \\ \hline
G   &  28 &  19 &  17 &  12 &  11 \\
K   &  66 &  30 &  59 &  23 &  19 \\
M   &  74 &  88 &  98 &  44 &  33 \\ \hline
\multicolumn{6}{c}{\bf Pleiades} \\ \hline
G   &  41 &  82 &  41 &  31 &  20 \\
K   & 118 & 231 & 112 &  87 &  59 \\
M   &  52 & 139 &  65 &  47 &  31 \\ \hline
\multicolumn{6}{c}{\bf Hyades} \\ \hline
G   &  29 &   2 &  22 &   8 &   8 \\
K   &  71 &  20 &  54 &  29 &  26 \\
M   &  84 &  69 &  99 &  46 &  33 \\ \hline
\end{tabular}
\end{table}

\begin{figure*}
\begin{center}
\parbox{18cm}{
\parbox{9cm}{\resizebox{9cm}{!}{\includegraphics{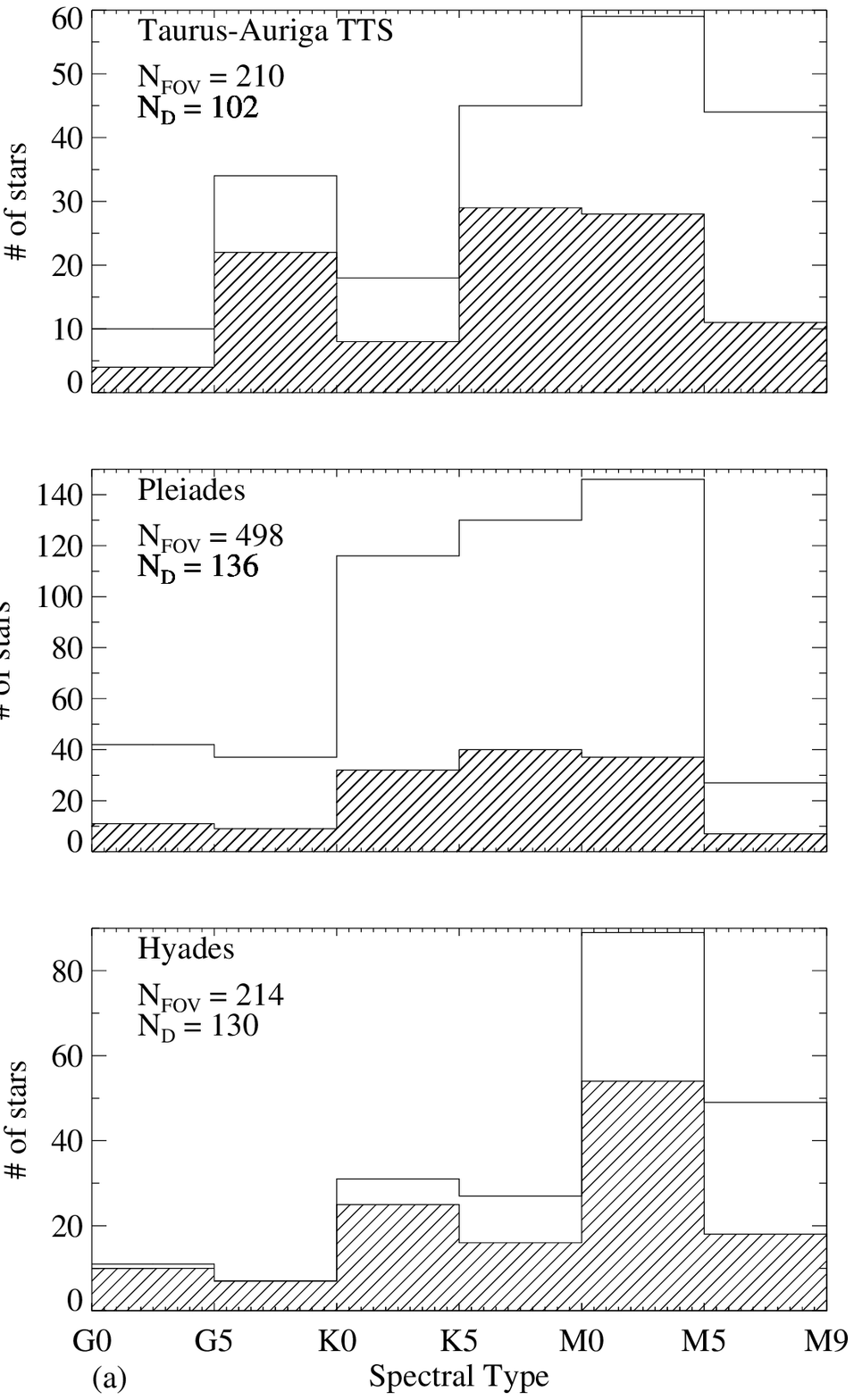}}}
\parbox{9cm}{\resizebox{9cm}{!}{\includegraphics{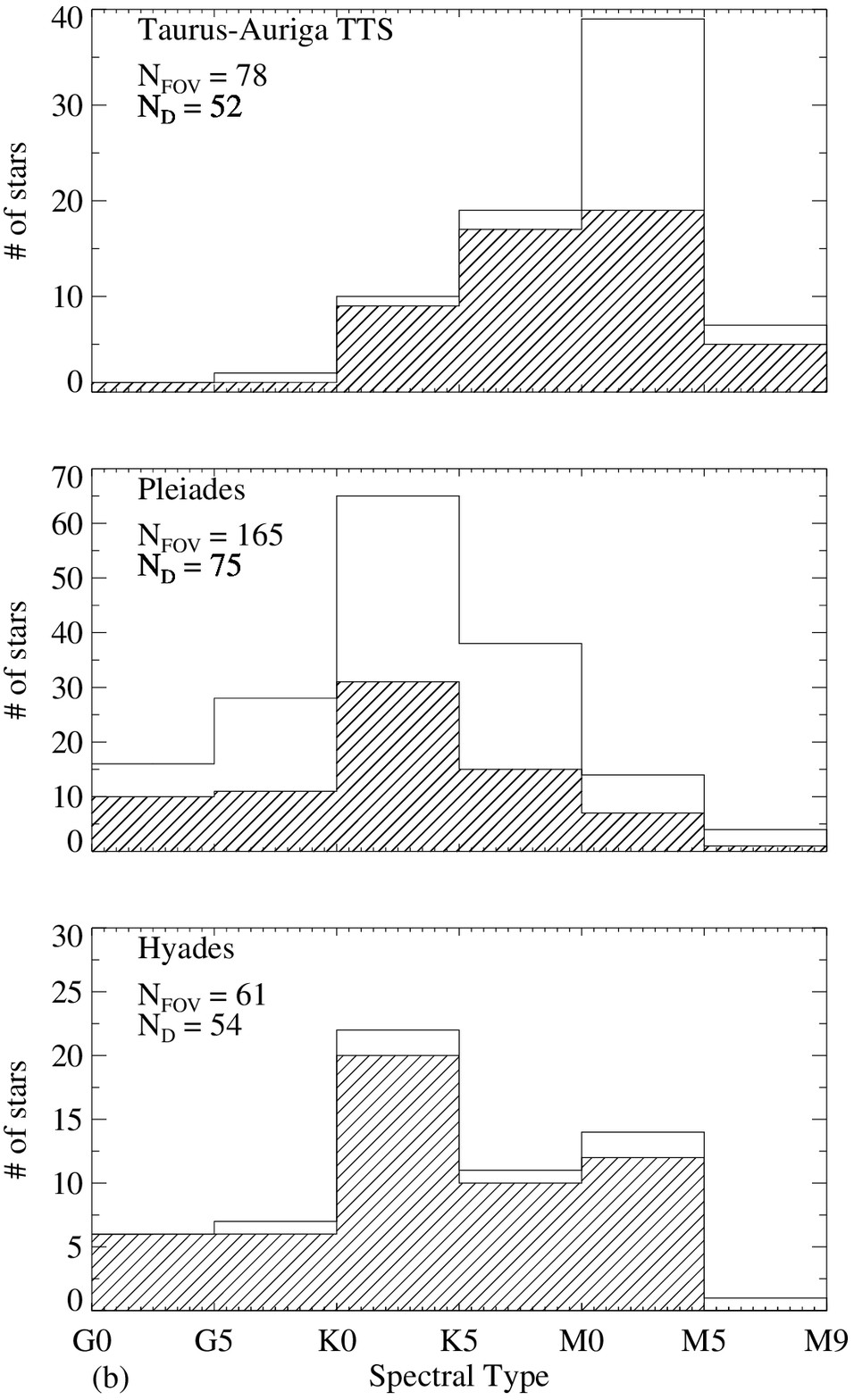}}}
}
\caption{Spectral type distribution of the observed late-type stars: (a) 
single stars, (b) multiple stellar systems. 
For multiples the spectral type of the primary is plotted, and the
secondary is not taken into account.
Solid lines 
denote the total number of stellar systems in any field of the {\em ROSAT} PSPC
observations from Table~1 in SNH00. The hatched
areas represent the number of detected systems. The numbers given in each 
panel represent the total number of observed systems (`$N_{\rm FOV}$') and
detected systems (`$N_{\rm D}$'). Note, that individual stellar systems may have
been detected in more than one observation. 
The fraction of detected stars depends
on distance, integration time, possible flaring activity, line-of-sight
absorption, and stellar parameters such as age, mass, and rotation.}
\label{fig:sptypes_histo}
\end{center}
\end{figure*}

Histograms of the distribution of spectral types in the different 
stellar samples are displayed in Fig.~\ref{fig:sptypes_histo}. 
We show separate histograms for single and multiple star systems.
In the latter sample only the primary is considered. (For most secondaries
the spectral type is unknown).
The empty histogram bins give 
the number of stars in the {\em ROSAT} PSPC field of any observation
and the hatched histograms the subgroup of detected stars. 
The total number of stars displayed in the figure is also given and labeled
`$N_{\rm FOV}$' (all observed stars) and `$N_{\rm D}$' (all detected stars)
respectively. 

As seen in Fig.~\ref{fig:sptypes_histo}, the detection rate is higher for
the Hyades than for the Pleiades or TTS, although the Hyades
are older.
This is probably due to their shorter distance. The relative number
of detections is larger for TTS than for the Pleiades presumably because
TTS are young and more active. 
Throughout all spectral types the detection rate is higher for unresolved 
binaries as compared to single stars. 
This could indicate that all stars in multiple systems
contribute to the X-ray emission.
The actual detection rate is a complicated
function of many influencing factors, such as distance, integration time,
absorption, age and mass. A detailed analysis of the X-ray
emission levels of the different groups of stars is given in the
following sections.

\subsection{X-ray Detection of Very-late Type Dwarfs}\label{subsect:m9dwarf}

A number of very low-mass dwarfs with spectral types between M5 and M9
have been detected. In particular, we report on the detection of
LH\,0429+17, to date the latest PMS dwarf with X-ray emission.
This object was listed as a candidate member of the Hyades in a
photometric study by \citey{Leggett89.1}. In the course of a spectroscopic
survey for brown dwarfs in the Hyades \citey{Reid99.1} have detected strong
H$\alpha$ emission but weak absorption in the gravity sensitive Na~I line, 
which is an indication for young age. Taking into account its location 
on the sky, LH\,0429+17 can, therefore, 
be considered as member of the Taurus star forming region.

X-rays from young brown dwarfs and 
brown dwarf candidates in the Chamaeleon, Taurus-Auriga and $\rho$\,Ophiuchus 
star forming regions have first been observed by \citey{Neuhaeuser98.1} and 
\citey{Neuhaeuser99.1}. 
These objects have spectral types between M6 and M8. 
Note, that we confirm here the detection of all brown dwarfs and
brown dwarf candidates in the Taurus region which have been listed in 
\citey{Neuhaeuser99.1}.

\section{H-R Diagrams}\label{sect:hrds}

\begin{figure}[h]
\begin{center}
\resizebox{9cm}{!}{\includegraphics{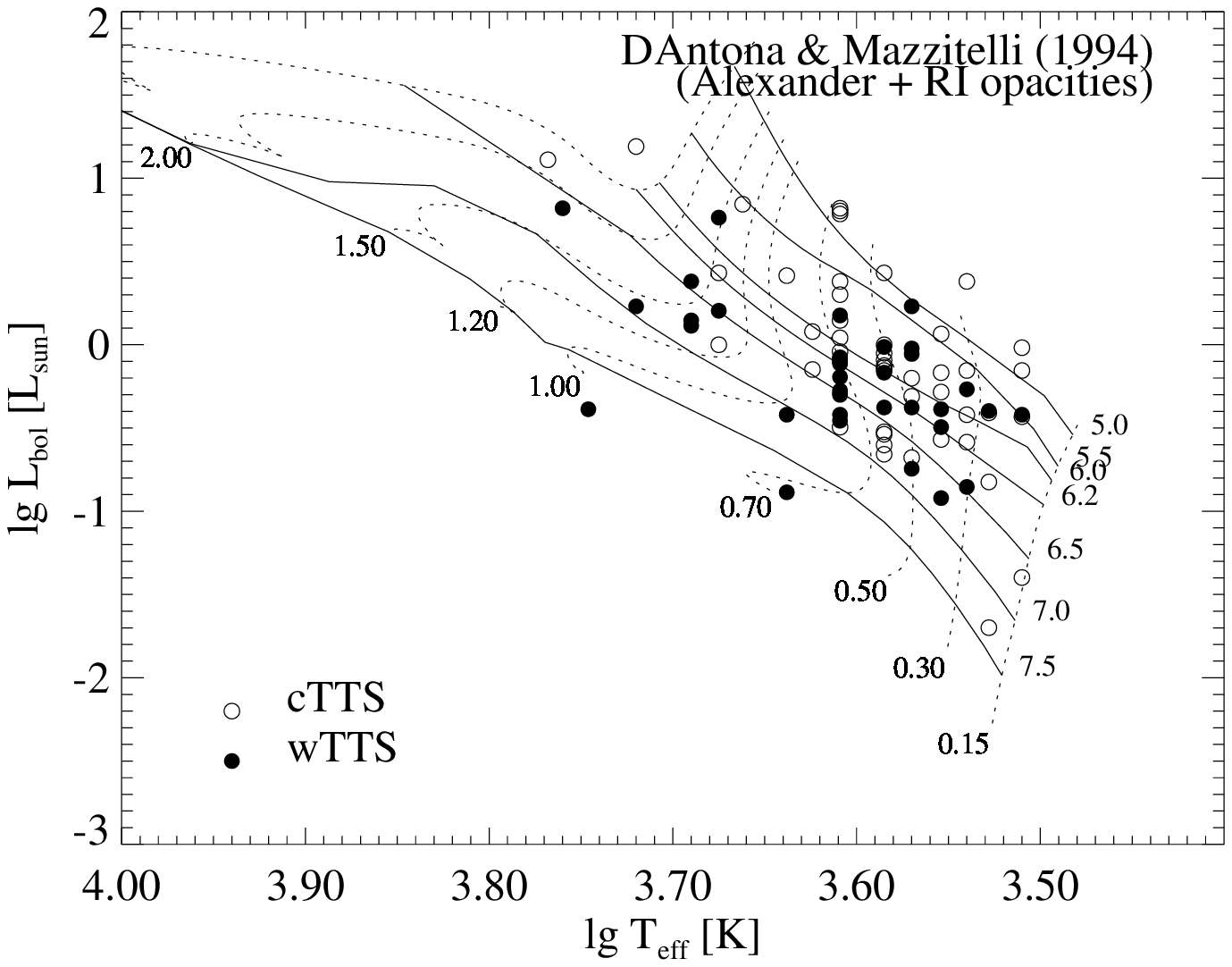}}
\resizebox{9cm}{!}{\includegraphics{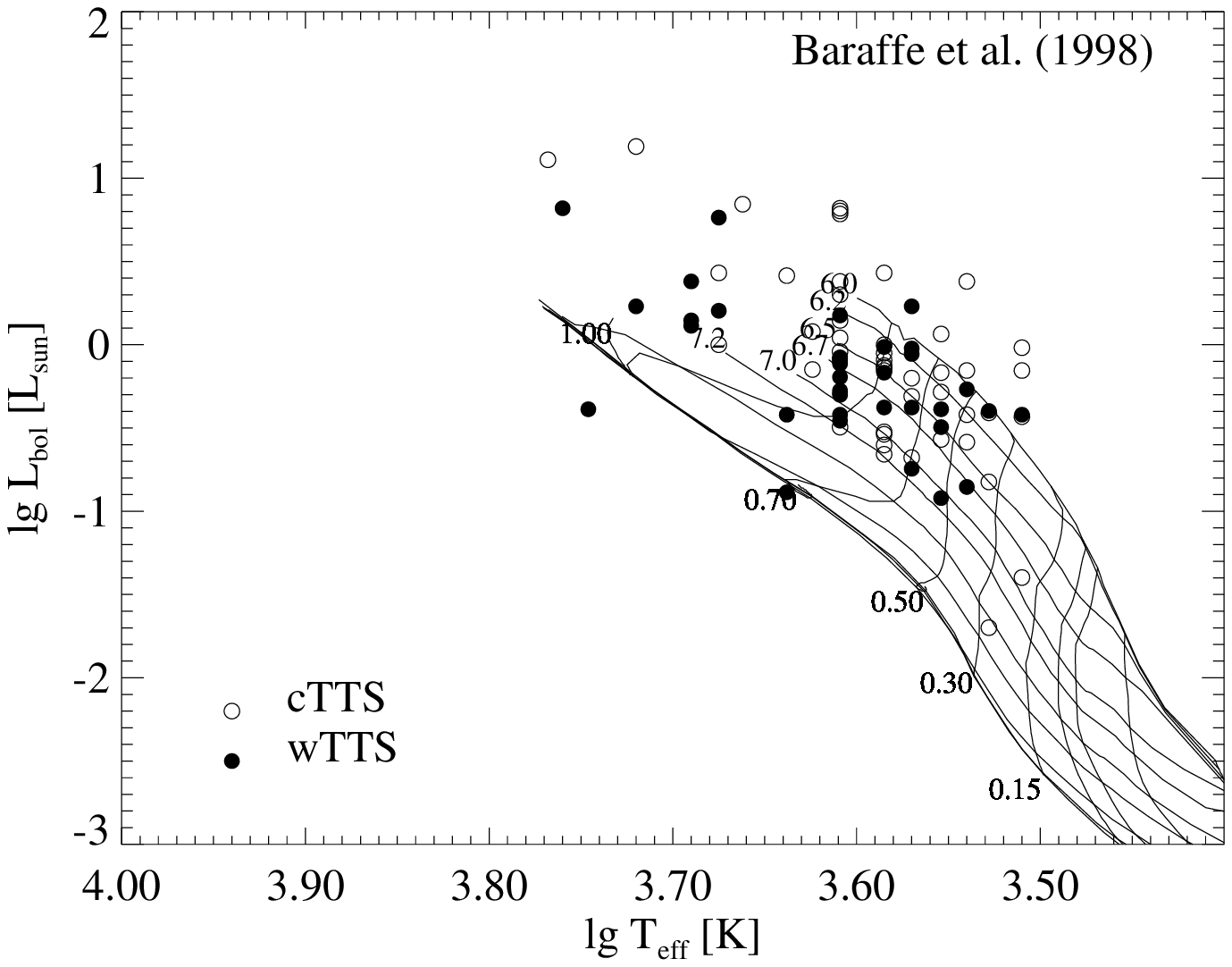}}
\resizebox{9cm}{!}{\includegraphics{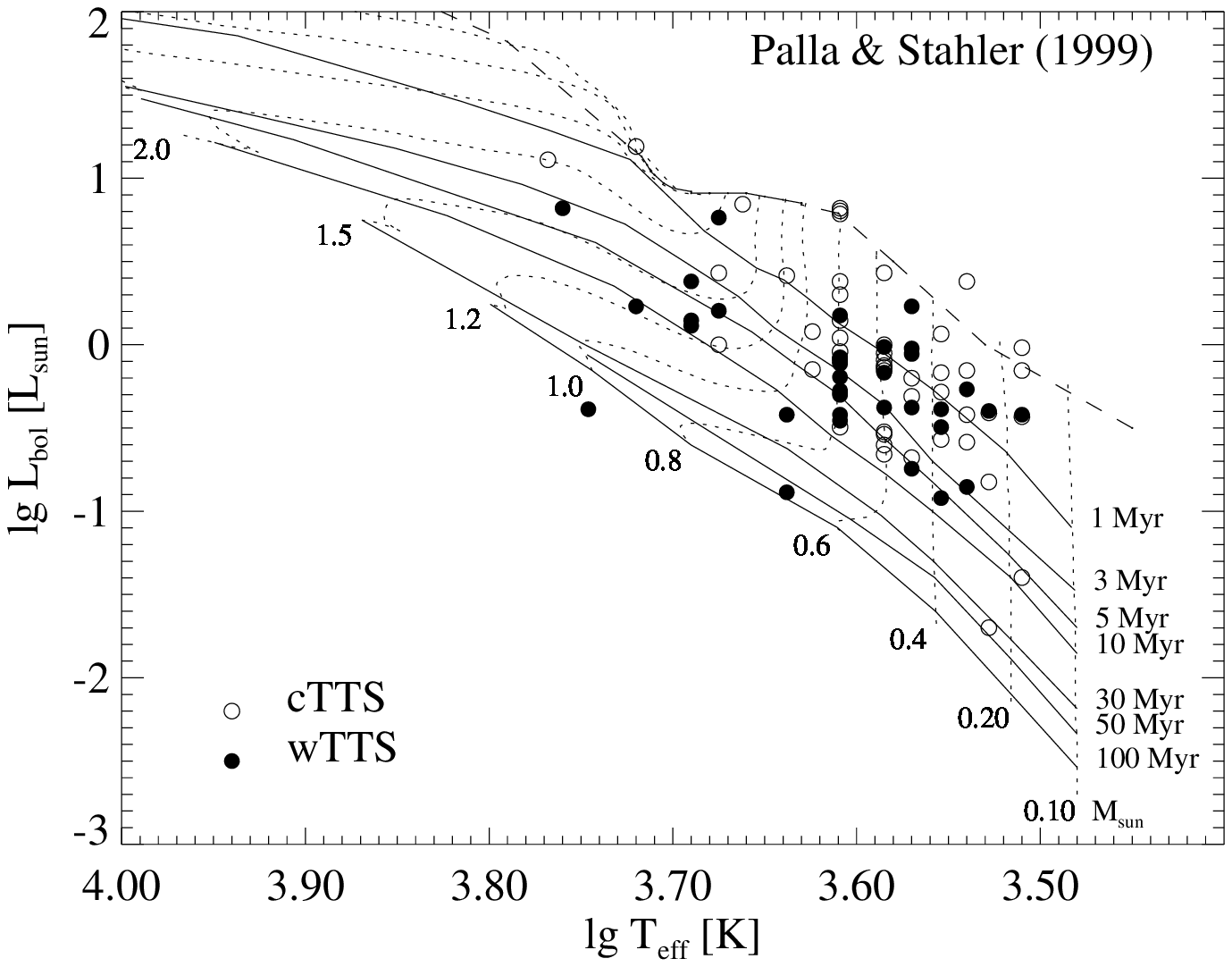}}
\caption{HR diagram of TTS observed with the {\em ROSAT}
PSPC during pointed observations. Note, that the stars on display represent
only a fraction of all X-ray observed TTS because $L_{\rm bol}$ and $T_{\rm
eff}$ are not known in all cases. The data are compared to three different
theoretical calculations for the PMS evolution: {\em top} -
\protect\citey{dAntona94.1}, {\em middle} - \protect\citey{Baraffe98.1},
and {\em bottom} - \protect\citey{Palla99.1}. The masses are given in solar
units and the isochrones represent $\lg{\rm age}$ except for
\protect\citey{Palla99.1} where the ages are given in Myrs.}
\label{fig:hrd}
\end{center}
\end{figure}
In order to visualize the age and mass distribution of our stars we have
placed the subset of TTS observed with the PSPC and 
with known bolometric luminosity $L_{\rm bol}$ and 
effective temperature $T_{\rm eff}$ in the 
Hertzsprung-Russell diagram (H-R diagram). We dispense with H-R diagrams  
for Pleiades and Hyades stars because most of the stars in these two
clusters are well known to lie on the MS (see previous discussion). 
The H-R diagram for the TTS in Taurus-Auriga is shown in Fig.~\ref{fig:hrd}.
We have used the $L_{\rm bol}$ values given by \citey{Kenyon95.1}, 
and $T_{\rm eff}$ was obtained from the spectral types 
using the conversion by \citey{Schmidt-Kaler82.1}.
The location of the stars is compared to different models of 
evolutionary PMS tracks: \citey{dAntona94.1},
\citey{Baraffe98.1}, and \citey{Palla99.1}. All diagrams are drawn with
the same scale to facilitate the perception of differences 
between the model calculations. The computation by \citey{Baraffe98.1}
does not represent a useful description of the complete TTS sample 
due to its restriction to masses below $\sim 1\,{\rm M_\odot}$.
Furthermore, the lines of equal mass show significant deviations from the
other calculations. A closer look reveals that there are also significant
differences between the models of \citey{dAntona94.1} and \citey{Palla99.1}.

It would be highly desirable to use 
the theoretical calculations to assign ages and masses to the individual TTS. 
However, from the comparison provided in Fig.~\ref{fig:hrd} it is obvious 
that the calibration of the models is uncertain, 
i.e. tracks computed by different groups would lead to controversial results
on the masses and ages of the stars.

Despite such uncertainties 
the H-R diagram can be used to demonstrate the average distribution
of the cTTS and wTTS. Although the stars closest to the birthline tend to 
be cTTS, and those nearest to the MS are wTTS, the overall
distribution of cTTS and wTTS is mixed.
This indicates that individual wTTS are not always older than cTTS 
despite the fact that they represent a later evolutionary stage.
This is known since the discovery of many wTTS by the {\em EO} 
(\cite{Walter88.1}), and the situation is similar in other star forming
regions.

\section{Luminosity Functions}\label{sect:ldfs}

The statistical analysis was performed with the ASURV package
version 1.2 (see \cite{Feigelson85.1}, \cite{Isobe86.1}, 
\cite{LaValley92.1}). The ASURV software is particularly well suited for 
the study of data sets with censored points, i.e. non-detections. We exclude 
photons observed during the large X-ray flares presented by SNH00, i.e. for
flaring stars only their quiescent radiation is taken into account.

XLF are frequently employed to characterize a stellar population.
Our special interest here is to compare the XLF of the different
stellar groups with respect to the following issues: (i) Are the
luminosity functions of cTTS and wTTS different, (ii) how does
the X-ray luminosity evolve with stellar age, (iii) how does
it depend on the spectral types of the stars and their binary character.

A substantial number of stars are in the field of more than one pointed PSPC
observation (see Table~\ref{tab:det}). 
However, every star should appear only once in the XLF. 
Therefore, we represent each star by its error weighted 
mean luminosity from all observations in which it was detected. 
If a star was observed in more than one observation, but not detected in
any of them, we use the mean upper limit of all non-detections of this
star as an estimate for its luminosity limit.

In 
Sect.~\ref{subsect:ldf_sing_bin} we will justify
our assumption 
that the X-ray luminosity can be distributed equally
among all stars in unresolved multiple systems. 
Therefore, if not specified otherwise, we have divided the mean X-ray luminosity
by the number of components in the stellar system.

\subsection{cTTS and wTTS in RASS and Pointed PSPC Data}\label{subsect:ldf_rass_point}

When studying the X-ray emission of TTS in Taurus-Auriga observed during
the RASS, N95 found that the wTTS are X-ray brighter than the cTTS. This is in
contrast to findings in various other star forming regions (see e.g. 
\cite{Feigelson93.1}, \cite{Casanova95.1}, \cite{Grosso00.1}). 
This discrepancy is not yet understood. 
A possible explanation is that the XLF of the wTTS in Taurus-Auriga 
is uncomplete towards the low-luminosity end, because
wTTS are not easily identified due to the lack of pronounced 
spectral features. 
In particular, many wTTS have been discovered with the {\em EO}.
Therefore, even the pre-{\em ROSAT} sample studied in N95 
could be biased towards X-ray bright wTTS.

\begin{figure}
\begin{center}
\resizebox{9cm}{!}{\includegraphics{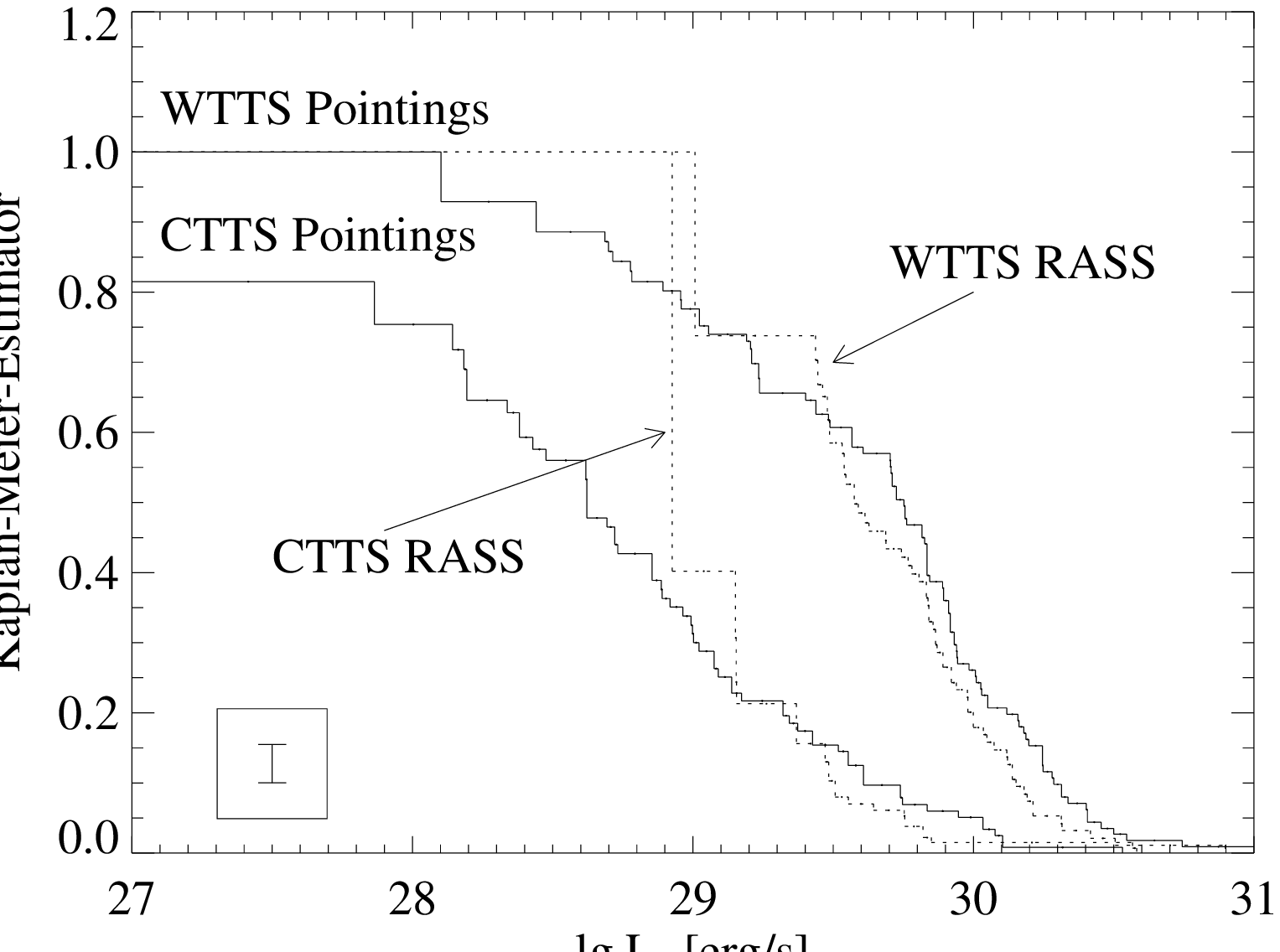}}
\caption{XLF of TTS in Taurus-Auriga derived from the
RASS and from pointed {\em ROSAT} PSPC
observations. Shown are all cTTS and wTTS in
Taurus-Auriga. The inset in the lower left shows the typical error bar.}
\label{fig:ldf_rass_point}
\end{center}
\end{figure}
Our analysis of a large set of pointed {\em ROSAT} observations allows
to extend the sensitivity limit substantially with respect to the RASS.
In Fig.~\ref{fig:ldf_rass_point} we compare the XLF of TTS derived 
from the pointed observations described in this paper to the results 
from the RASS.
The XLF of the RASS have been newly compiled with respect to the analysis
by N95 to include all TTS discovered since then, i.e. the sample consists 
of all TTS from N95 plus those listed in \citey{Koenig01.1} 
(including both {\em EO} and {\em ROSAT} discovered
TTS). N95 did include {\em EO} discovered but no
{\em ROSAT} discovered TTS. 

The comparison with the RASS data clearly demonstrates 
the better sensitivity of
the pointed observations. The XLF computed from the PSPC pointings extends
by $\sim 1-2$ orders of magnitude further into the low luminosity 
regime. We reproduce the result first found by 
N95: In Taurus-Auriga the wTTS are clearly X-ray brighter
than the cTTS. 

\begin{figure}
\begin{center}
\resizebox{9cm}{!}{\includegraphics{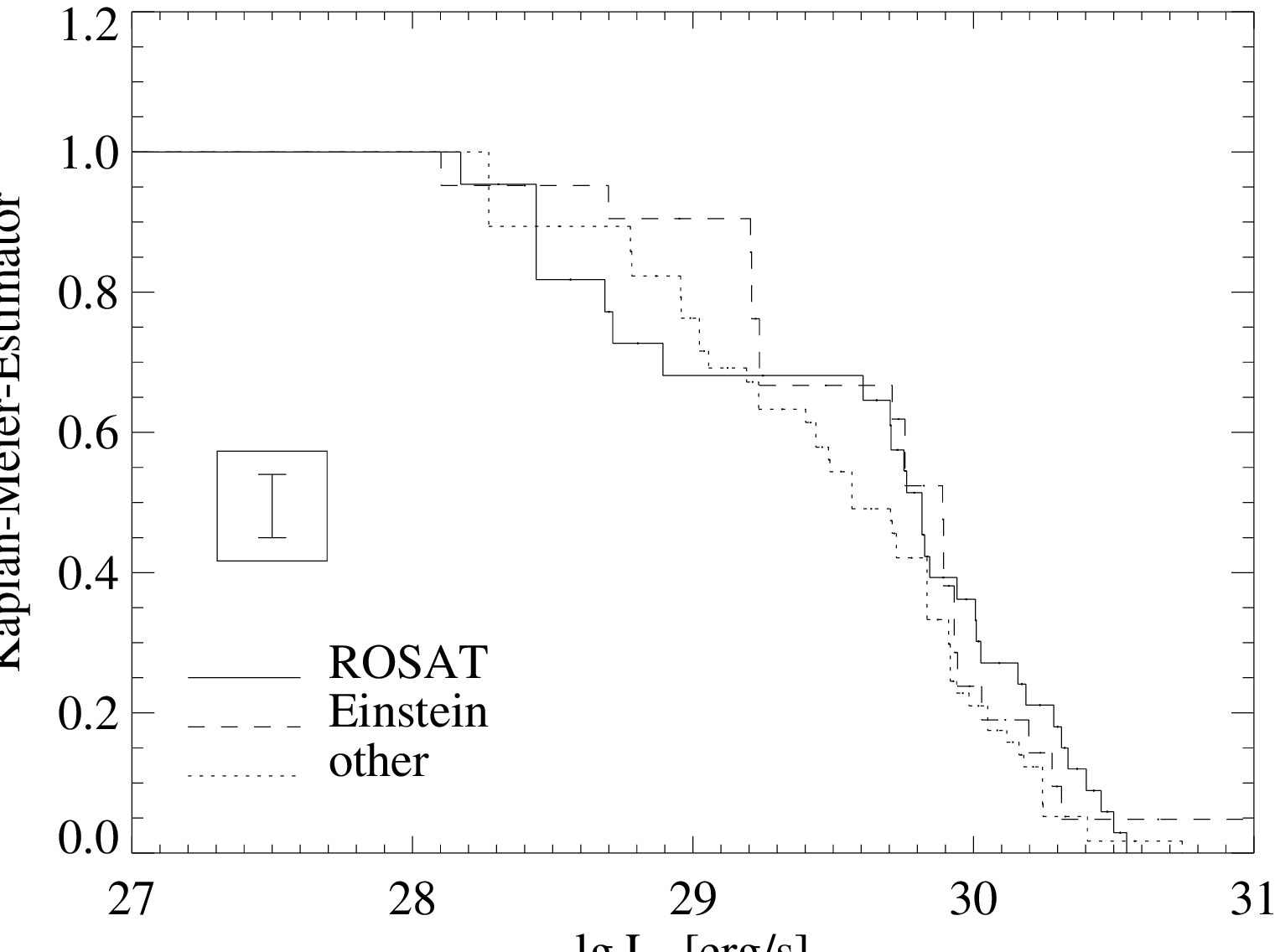}}
\caption{XLF of wTTS in Taurus-Auriga derived 
from pointed {\em ROSAT} PSPC observations. The three different
distributions are {\em ROSAT} discovered wTTS (solid line), {\em
EO} discovered wTTS (dashed line), and wTTS discovered by other
means (dotted line). 
The inset in the lower left shows the typical error bar.
All distributions are similar indicating that the
inclusion of X-ray discovered TTS does not introduce a selection bias into
the sample of wTTS.}
\label{fig:ldf_xdis}
\end{center}
\end{figure}

It was noted by \citey{Feigelson93.1} that the 
XLF can change, if the stars
included in the sample were found by different methods, 
e.g. H$\alpha$ versus X-ray surveys.
In order to overcome this bias we have computed XLF where we exclude 
all X-ray discovered TTS. Fig.~\ref{fig:ldf_xdis} shows the Kaplan-Meier
Estimator (KME) for 
three subsets of wTTS in Taurus-Auriga: {\em ROSAT} discovered wTTS, 
{\em EO} discovered wTTS, and all other wTTS.
The XLF of these groups do not differ significantly from each other.
Therefore, we are led to conclude that the difference in the distributions 
of cTTS and wTTS is not due to an X-ray selection bias.

\begin{figure}
\begin{center}
\resizebox{9cm}{!}{\includegraphics{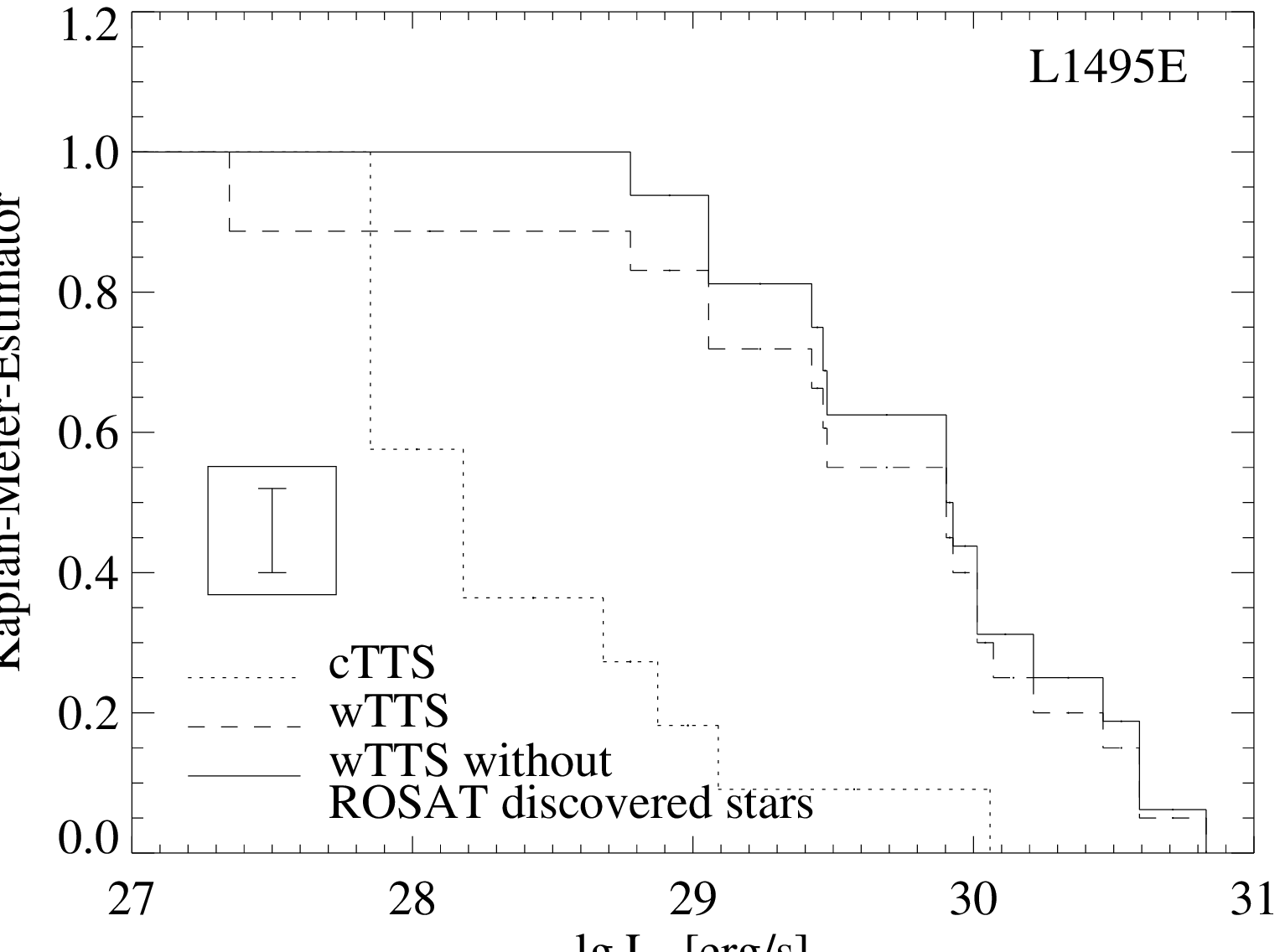}}
\caption{XLF of wTTS and cTTS in L1495E derived 
from a $\sim$\,30\,ksec pointed {\em ROSAT} PSPC observation: {\em dotted
line} - cTTS; {\em dashed line} - all wTTS; {\em solid line} - wTTS except those discovered by {\em ROSAT}.
The inset in the lower left shows the typical error bar.
The distributions of cTTS and wTTS are again different indicating that the
discrepancy between the X-ray luminosities of cTTS and wTTS in
Taurus-Auriga is not due to the spatial extension of the sample.}
\label{fig:ldf_strom}
\end{center}
\end{figure}

The differences to the $\rho$\,Oph and Cha\,I star forming regions 
(\cite{Feigelson93.1}, \cite{Casanova95.1}, \cite{Grosso00.1})
could also be caused by the difference in spatial extension 
between these two young clusters and the Taurus-Auriga region:
The latter is widely dispersed, and, hence, its members may constitute 
a larger spread in age as compared to the more 
complex $\rho$\,Oph and Cha\,I regions
in which the stars are probably more coeval.
We can check this by selecting TTS from the central parts of the star
forming region,
and comparing the resulting XLF with that of the total sample.
We have chosen the PSPC observations ROR 200001-0p and 200001-1p
pointed on the L1495E cloud. These pointings
are centered on the largest concentration of molecular material
corresponding to a particular young part of the Taurus complex.
In Fig.~\ref{fig:ldf_strom}
we show the XLF for wTTS and cTTS in that region. A third distribution
consists of all wTTS in L1495E 
which have {\em not} been discovered by {\em ROSAT}.
The general shape of the XLF in L1495E is the same as that for the
complete Taurus-Auriga area: wTTS are X-ray brighter than
cTTS. This is also evident from the data in \citey{Strom94.1},
an earlier analysis of these pointings in L1495E.
We conclude that the X-ray luminosity does not depend on the spatial
location within the Taurus region. 

The difference in the XLF of wTTS and cTTS does also not depend on our
choice of roughly 10\,\AA~ as boundary between cTTS and wTTS.
It is clear that one should use the H$\alpha$ flux instead of the
equivalent width as boundary (hence, we classify SU\,Aur as cTTS)
because the equivalent width depends on the underlying continuum
which varies with spectral type.
\citey{Martin97.1} suggested three different equivalent width boundaries
for three spectral type regimes chosen such as to exclude that the H$\alpha$
emission is due to chromospheric activity.
Adopting these criteria only
a few TTS change classification, but the difference in the XLF remains.

In Sect.~\ref{subsect:res_soudet} the conversion from count rates to
luminosities by use of hardness ratios was explained. Using hardness
ratios allows to indirectly take account of the extinction in the
absence of actual $A_{\rm V}$ measurements. However, $HR1$ is only
sensitive to comparatively low extinctions. The extinction should
generally be higher for the cTTS than for the wTTS due to the denser
circumstellar environment of the former ones, and if not treated
properly may lead to wrong estimates for the luminosities. 

We have, therefore,
applied an alternative way of deriving X-ray luminosities for the 
TTS in Taurus-Auriga making use of the available $A_{\rm V}$ data. 
In this approach the X-ray flux
was computed with standard EXSAS tools assuming a 1\,keV RS-model with
absorbing column density $N_{\rm H}$ derived from $A_{\rm V}$ according to 
\citey{Paresce84.1}. Similar values for $N_{\rm H}$ are obtained when 
using the conversion given by \citey{Ryter96.1}. 
Stars for which $A_{\rm V}$ is $\leq 0.05$\,mag  
have been assigned a standard value of $N_{\rm H} = 10^{18}\,{\rm cm}^{-2}$.

While for individual stars the $L_{\rm x}$ derived by
the two methods show typical deviations of $\sim 50$\,\%, 
the statistical distribution of X-ray luminosities is unaffected
by the specific choice of $CECF$,  
and the previously
discussed differences between the XLF of cTTS and wTTS remain.

\subsection{Dependence on Spectral Type}\label{subsect:ldf_sptypes}

In the previous subsection, no distinction was drawn between stars of 
different spectral type, mass or other 
stellar parameters. This is justified for young, 
very low-mass stars which follow fully convective tracks. 
It is believed that for stars on the MS
activity is governed by the relative size of radiative core and 
convective envelope. This should also apply to TTS once they have reached
the radiative part of their PMS evolution. Therefore, to obtain homogeneous
samples, stars with different interior structure, i.e. different mass,
should be treated separately. 
As argued in Sect.~\ref{sect:hrds} it is not possible to obtain
reliable values for the individual masses and ages of the stars. As an
approximation we distinguish the stars by their 
spectral type. But note, that while for stars on their Hayashi tracks
this description is acceptable, 
for stars on radiative tracks a given spectral type represents a mass range
rather than a single value for the mass.

Each subsample is subdivided in three spectral type bins: G, K, and 
M stars.
The mean X-ray luminosities 
for the different stellar groups and spectral
types are listed in Table~\ref{tab:lxmean}. For all spectral types the
wTTS distribution shows the largest values of 
$\langle \lg{L_{\rm x}} \rangle$, 
and the Hyades have the lowest $\langle \lg{L_{\rm x}} \rangle$. 
Note, that 
the group of cTTS of spectral type G is represented by only two stars. 
But for the other subsamples the statistics are significant.
Since in most cases the spectral type (or $B-V$) is known only for the
primary in multiples, we exclude the secondaries from this part of the
analysis, except the few cases where the spectral types of all components
are known (see Tables~2~-~7).

\begin{figure}
\begin{center}
\resizebox{9cm}{!}{\includegraphics{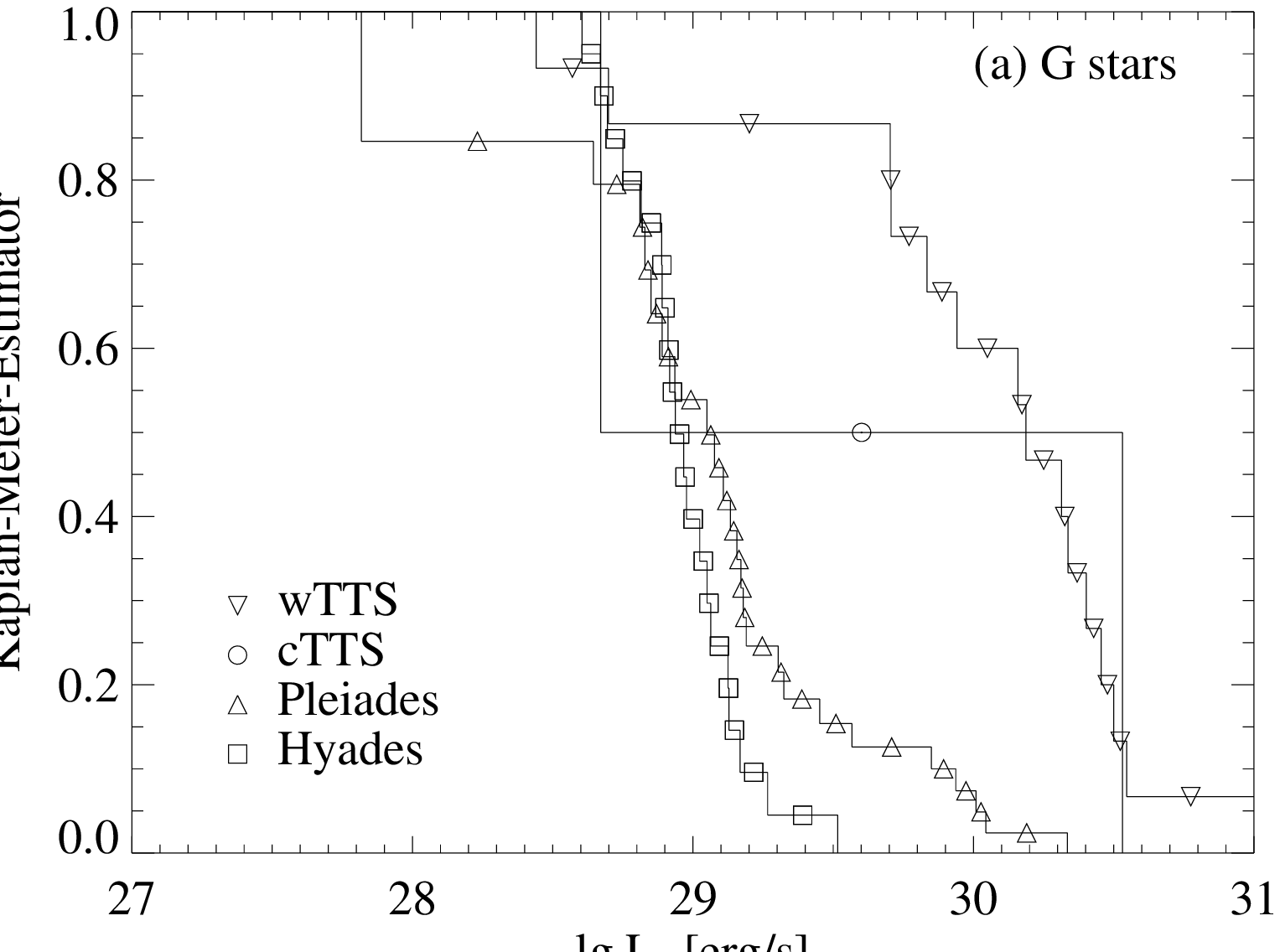}}
\resizebox{9cm}{!}{\includegraphics{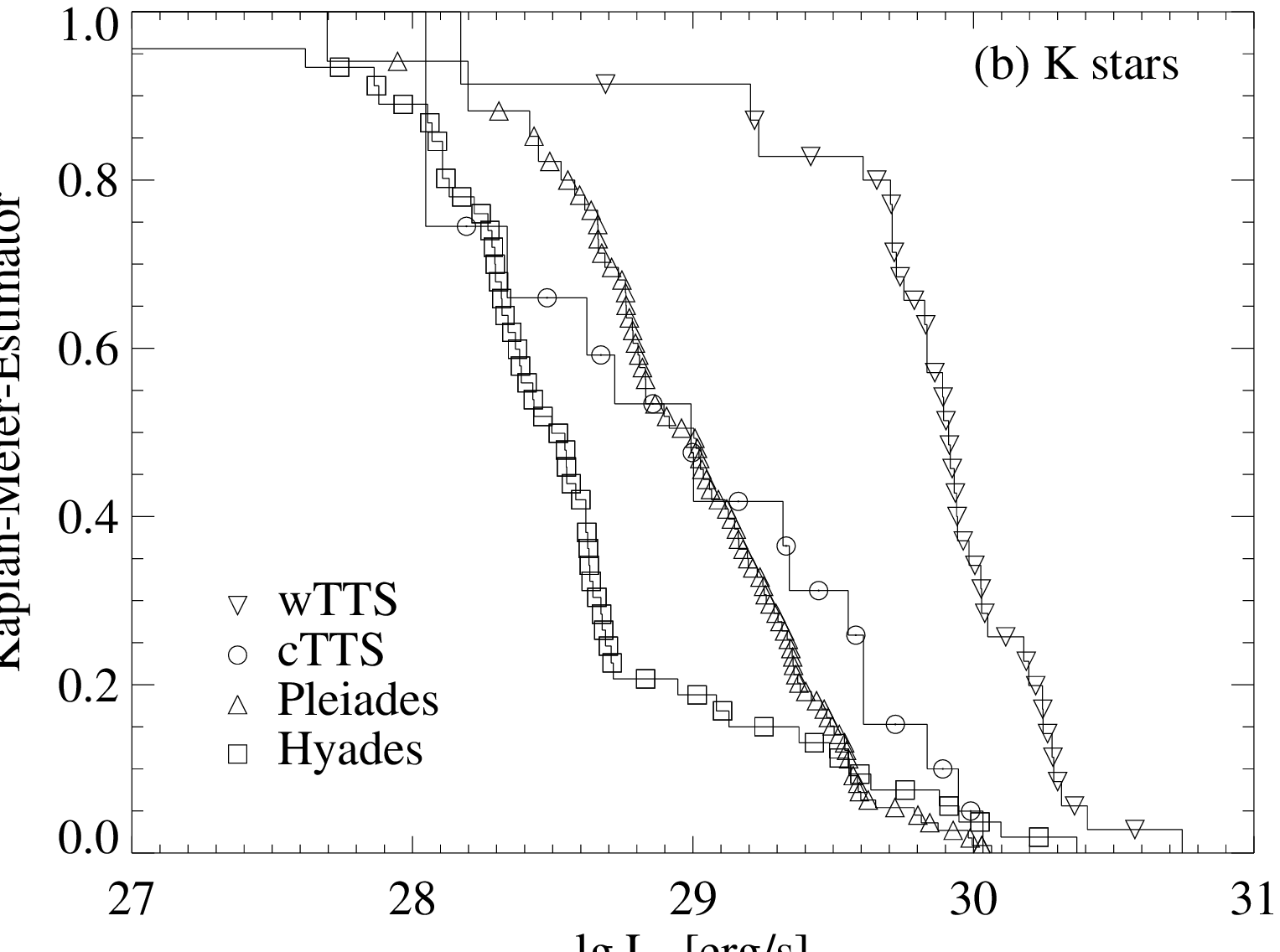}}
\resizebox{9cm}{!}{\includegraphics{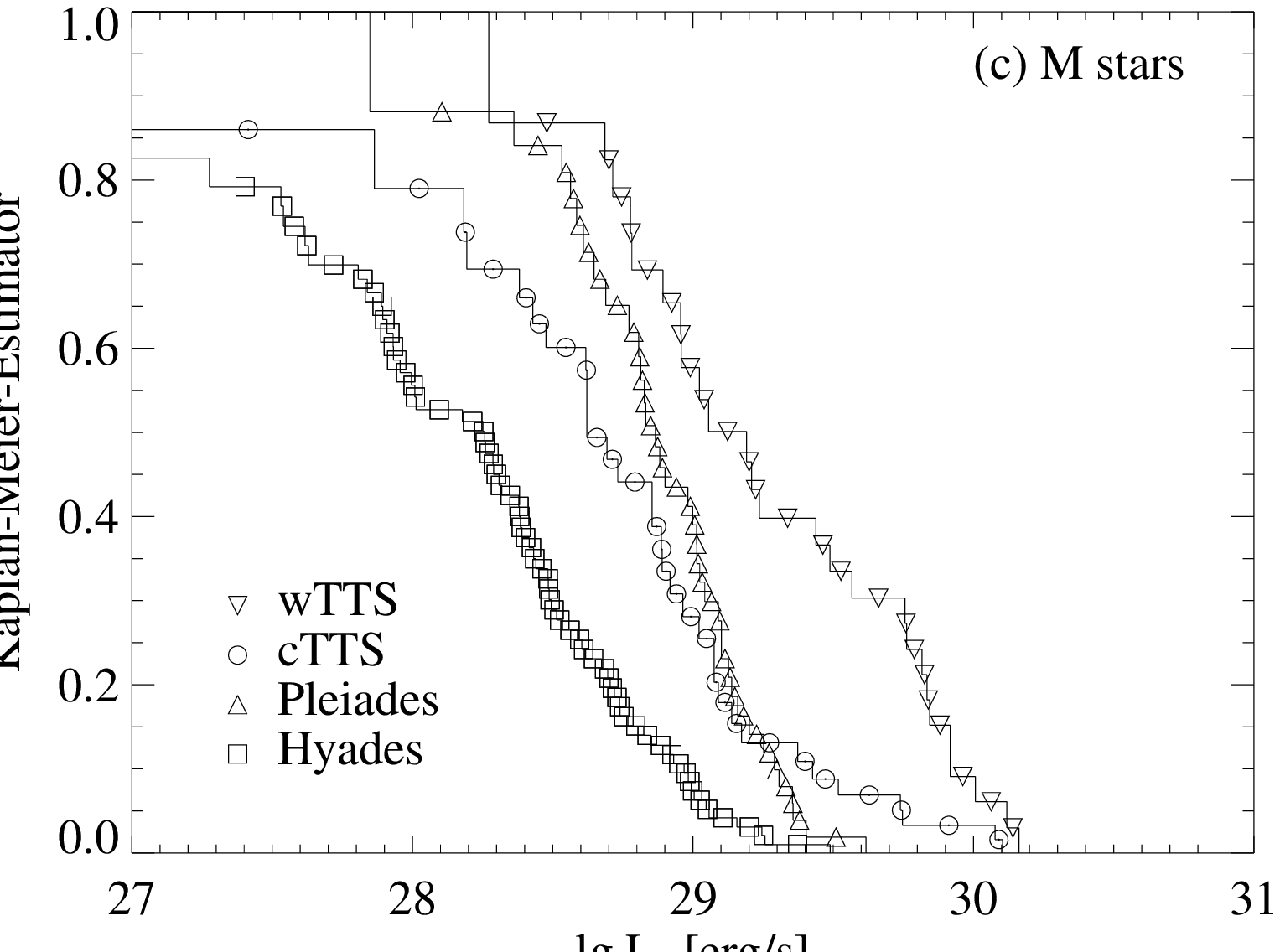}}
\caption{XLF for TTS in Taurus-Auriga, for the Pleiades, and the
Hyades. The distributions are shown for different spectral types,
corresponding to different values of $B-V$ or effective temperature or mass
for the MS stars. (a) G stars, (b) K stars, and (c) M stars.}
\label{fig:ldf_sptypes}
\end{center}
\end{figure}

In Fig.~\ref{fig:ldf_sptypes} we provide a comparison of the XLF of
TTS, Pleiads, and Hyads. 
Throughout all examined spectral types the 
wTTS clearly represent the brightest class among the X-ray objects 
studied here, and Hyads show the weakest X-ray emission.
For the M stars, where the mass range is comparatively small,
the decline of $L_{\rm x}$ from TTS over Pleiades to the Hyades can
be interpreted as an age effect. G and K type stars represent a
larger spread in the mass distribution such that the influences of 
mass and age 
may not easily be disentangled. However, the difference between
$\langle L_{\rm x} \rangle$ 
of Pleiades and Hyades decreases towards earlier types indicating
that age and not mass is the dominant effect.

The distributions of cTTS and
Pleiads intersect each other because of the much
shallower slope of the XLF of cTTS, i.e. larger spread in luminosities.
This effect may be caused by our assumption of uniform distance for
all stars in a given sample: In contrast to the strongly clustered Pleiades
region the TTS in Taurus-Auriga may be subject to a larger distance
spread that translates to an apparent spread in $L_{\rm x}$.

%
\begin{table*}
\caption{Mean X-ray luminosities $\langle \lg{L_{\rm x}} \rangle $
for cTTS, wTTS, and Pleiades and
Hyades. The columns labeled `$N$' and `$N_{\rm lim}$' give the number
of stars and number of upper limits within the sample. The second column
provides a description of the sample: 'C' - cTTS, 'W' - wTTS, 's' - single
star, 'b1' - binary star assuming that all X-ray emission comes from one
component, 'b2' - binary star assuming equal X-ray emission from both components.}
\label{tab:lxmean}
\begin{center}
\begin{tabular}{lcrrcrrcrrc} \hline
Region &  & \multicolumn{3}{c}{Spectral Type G} & \multicolumn{3}{c}{Spectral Type K} & \multicolumn{3}{c}{Spectral Type M} \\
& & $N$ & $N_{\rm lim}$ & $\lg{L_{\rm x}}$ & $N$ & $N_{\rm lim}$ & $\lg{L_{\rm x}}$ & $N$ & $N_{\rm lim}$
& $\lg{L_{\rm x}}$ \\ \hline
TTS & C & 2 & (1) & $29.60 \pm 0.66$ & 22 & (9) & $28.93 \pm 0.16$ & 61
& (30) & $28.54 \pm 0.14$ \\ 
TTS & W & 15 & (0) & $30.02 \pm 0.17$ & 36 & (5) & $29.78 \pm 0.10$ &
34 & (9) & $29.20 \pm 0.10$ \\
Pleiades & & 41 & (18) & $28.98 \pm 0.12$ & 112 & (41) & $28.94 \pm
0.06$ & 65 & (29) & $28.80 \pm 0.07$ \\ 
Hyades & & 22 & (2) & $28.97 \pm 0.05$ & 54 & (6) & $28.52 \pm 0.11$ & 99
& (38) & $27.99 \pm 0.10$ \\ \hline
TTS & s & - & - & - & 34 & (11) & $29.44 \pm 0.14$ &
60 & (28) & $28.70 \pm 0.14$ \\
TTS & b2 & - & - & - & 17 & (3) & $29.47 \pm 0.18$ &
29 & (20) & $28.85 \pm 0.18$ \\
TTS & b1 & - & - & - & 17 & (3) & $29.77 \pm 0.18$ & 29
& (10) & $29.15 \pm 0.18$ \\ 
Pleiades & s & 25 & (13) & $28.98 \pm 0.15$ & 84 & (38) & $28.83 \pm 0.09$ & 60
& (29) & $28.78 \pm 0.08$ \\ 
Pleiades & b2 & 16 & (5) & $29.03 \pm 0.16$ & 27 & (3) & $29.00 \pm 0.08$ & 5
& (0) & $28.93 \pm 0.08$ \\ 
Pleiades & b1 & 16 & (5) & $29.33 \pm 0.16$ & 27 & (3) & $29.30 \pm 0.08$ & 5
& (0) & $29.23 \pm 0.08$ \\ 
Hyades & s & 12 & (1) & $28.97 \pm 0.06$ & 36 & (5) & $28.41 \pm 0.15$ & 89 &
(35) & $27.95 \pm 0.10$ \\ 
Hyades & b2 & 10 & (1) & $28.96 \pm 0.07$ & 18 & (1) & $28.75 \pm 0.14$ & 9 &
(3) & $28.47 \pm 0.19$ \\ 
Hyades & b1 & 10 & (1) & $29.26 \pm 0.07$ & 18 & (1) & $29.05 \pm 0.14$ & 9 &
(3) & $28.77 \pm 0.19$ \\ \hline
\end{tabular}
\end{center}
\end{table*}

\begin{figure}
\begin{center}
\resizebox{9cm}{!}{\includegraphics{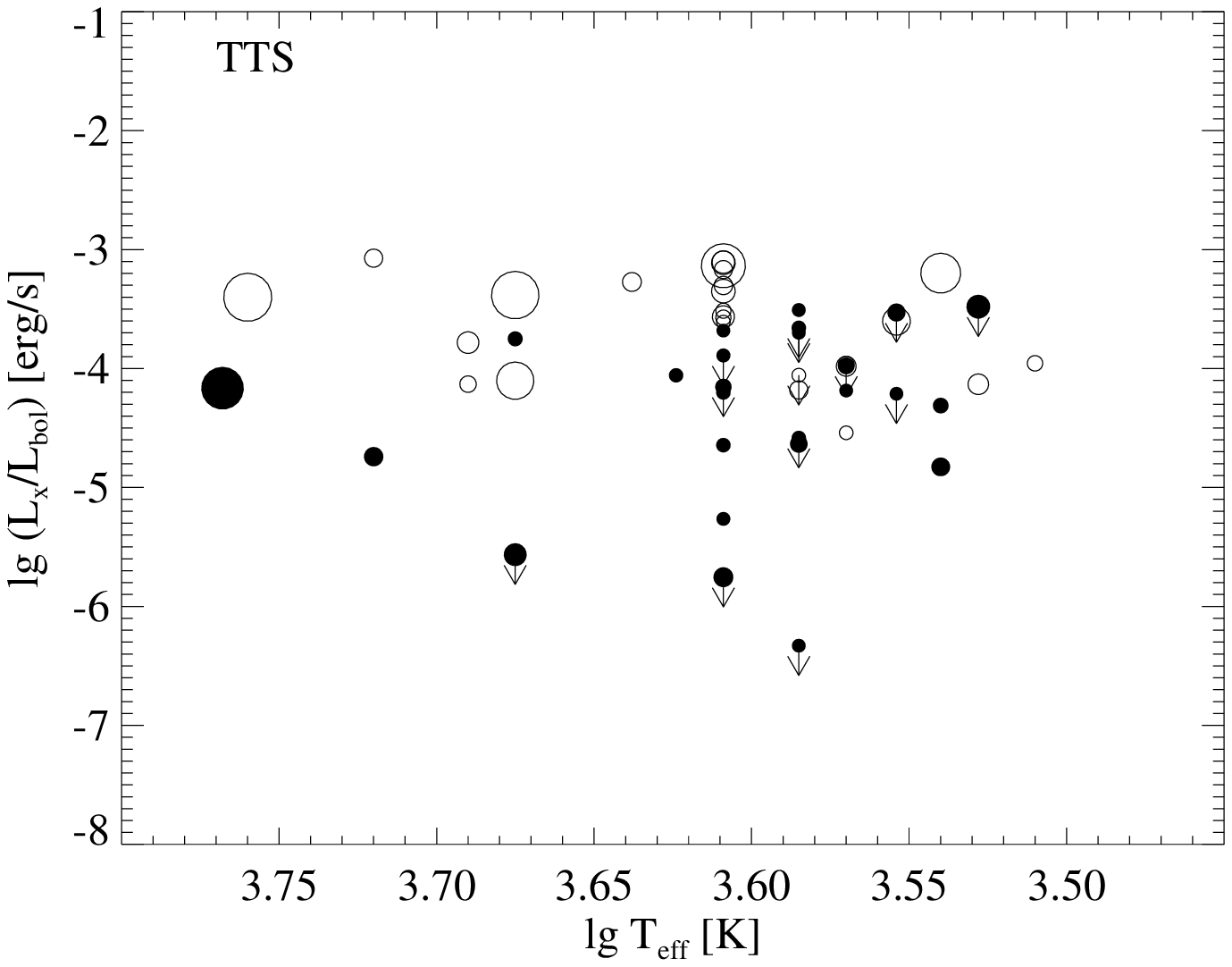}}	
\resizebox{9cm}{!}{\includegraphics{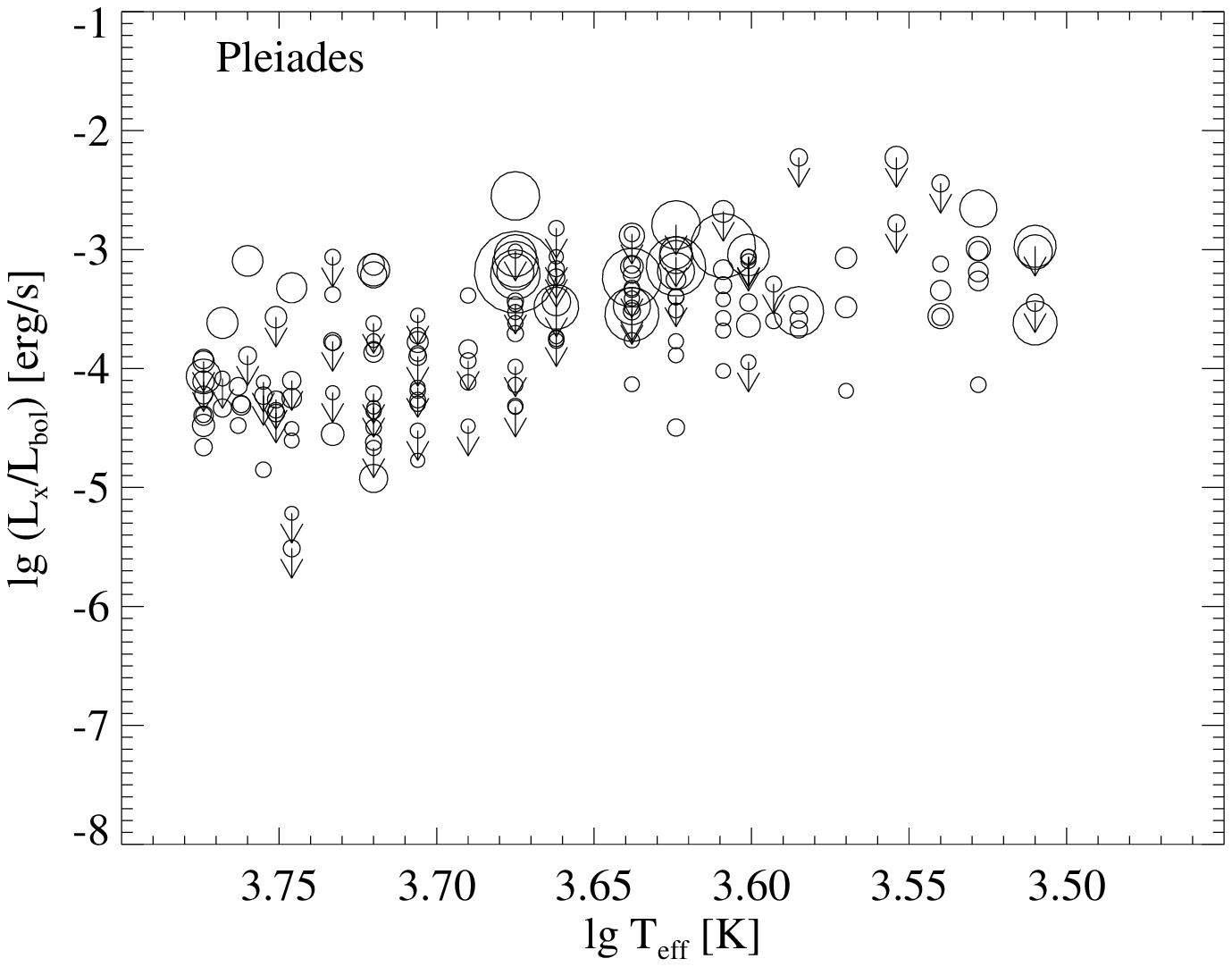}}	
\resizebox{9cm}{!}{\includegraphics{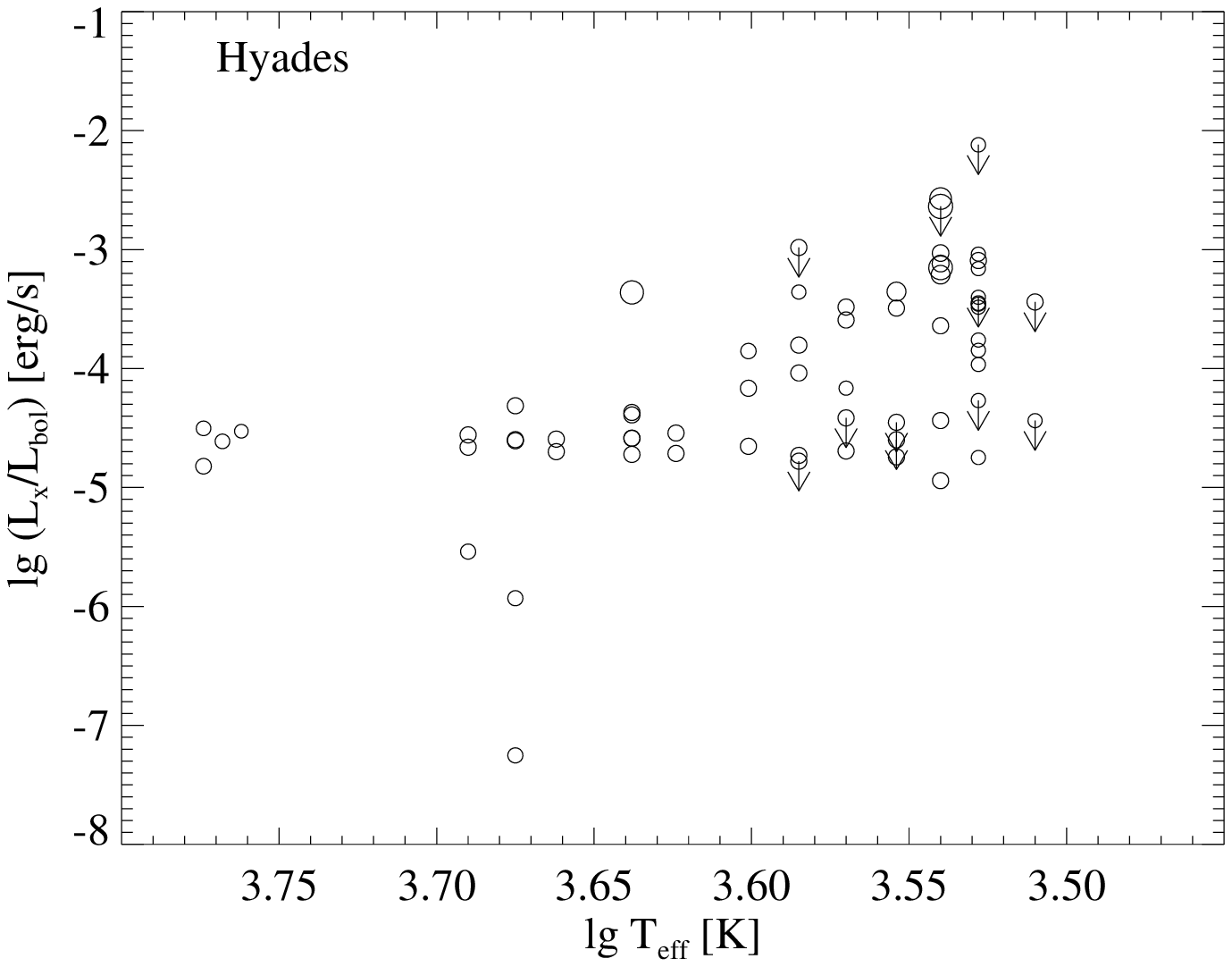}}	
\caption{Relation between X-ray to bolometric luminosity ratio, \lgLrat, 
and effective temperature, \lgTeff. From
top to bottom: TTS, Pleiades, and Hyades. Only stars with 
spectral type later than F are considered. The
plotting symbols have been scaled to the projected rotational velocity of
the stars. Upper limits to $L_{\rm x}$ are indicated by arrows.}
\label{fig:act_bv}
\end{center}
\end{figure}
Luminosity differences between various stars may generally be due to 
differences in emitting area. In order to eliminate this effect
the X-ray to bolometric luminosity ratio, \lgLrat, is often used to 
characterize the X-ray emission. 
We have examined the relation between the effective temperature
and \lgLrat. 
$L_{\rm bol}$ of Pleiads and Hyads was computed from
the $V$ magnitude and $B-V$ (needed to determine the bolometric correction)
given in the Open Cluster Data Base. 
The effective temperatures of Pleiades and Hyades stars were 
obtained from $B - V$. 
We have assumed negligible absorption to both star clusters.
In Fig.~\ref{fig:act_bv} all late-type stars 
(spectral type later than F or \lgTeff $< 3.78$) are plotted.
Fig.~\ref{fig:act_bv} shows that within the TTS sample, which is
characterized by a decline of 
$\lg{L_{\rm x}}$ with spectral type,
\lgLrat does not depend on effective temperature.
Pleiades and Hyades, however, demonstrate a clear anticorrelation between 
\lgLrat and \lgTeff (see also e.g. \cite{Micela99.1}). 
The fact that we do not see such a trend in the TTS sample may be due to the
large age spread among the TTS. 
Note, that in Fig.~\ref{fig:act_bv}
only stars with known projected rotational velocity are shown.
The plotting symbols have been scaled to \vsini. With few exceptions 
the fastest rotators are situated close to the upper envelope,
indicating a connection between the activity level and the
rotation rate (see also Sect.~\ref{sect:x_rot}).

\subsection{Single and Binary Stars}\label{subsect:ldf_sing_bin}

\begin{figure*}
\begin{center}
\parbox{18cm}{
\parbox{9cm}{\resizebox{9cm}{!}{\includegraphics{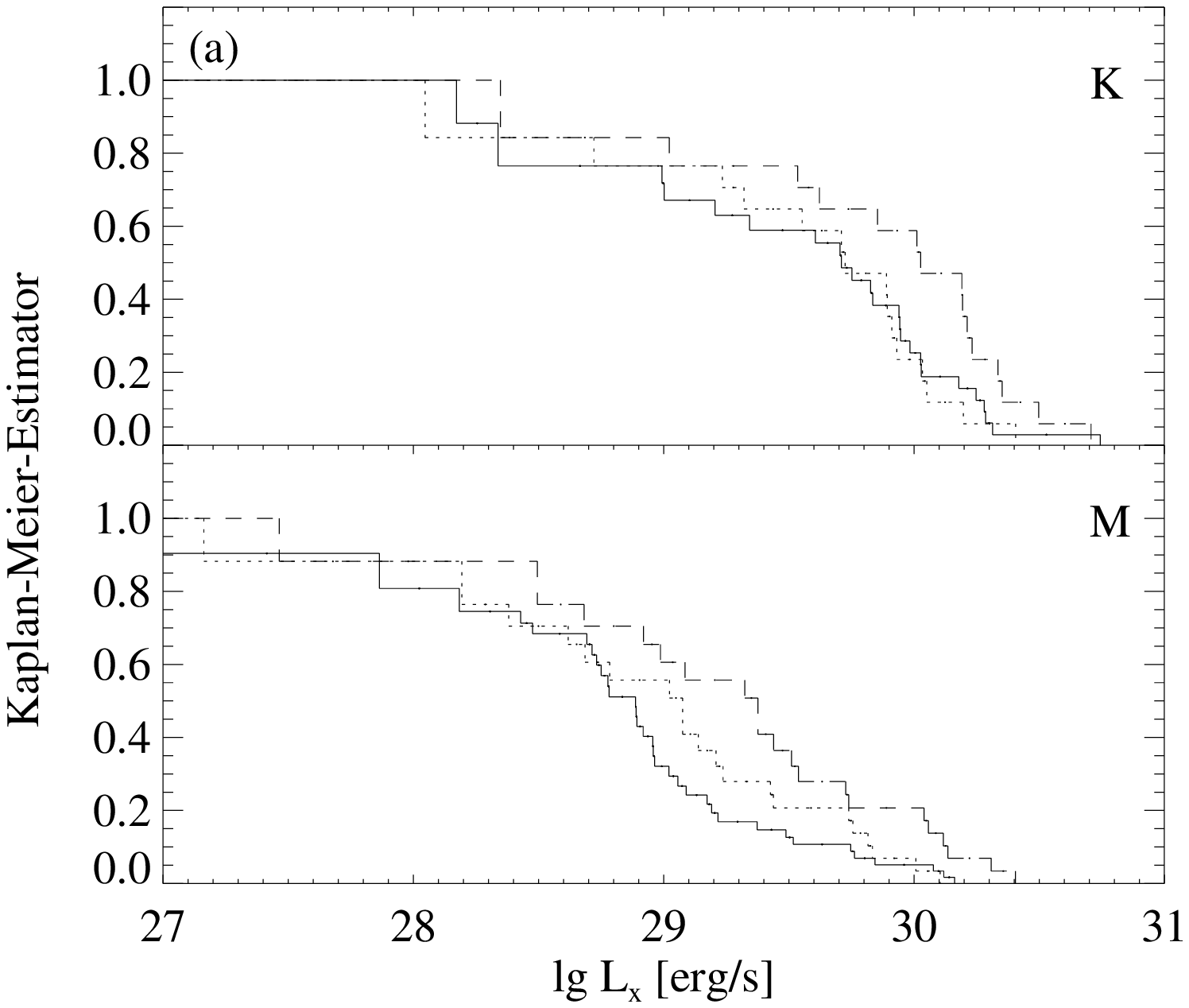}}}
\parbox{9cm}{\caption{XLF of single and binary stars of spectral type G, K
and M. (a) TTS, (b) Pleiades, and (c) Hyades. {\em solid lines} - single
stars (s), {\em dotted lines} - binary stars assuming 
equal $L_{\rm x}$ from both
components (b2), {\em dashed lines} - binary stars assuming only one X-ray
emitting component (b1). See text for a more detailed description of these
samples. All G type TTS in Taurus-Auriga are single stars, and therefore not displayed
in this figure.}\label{fig:ldf_sing_bin}}
}
\parbox{18cm}{
\parbox{9cm}{\resizebox{9cm}{!}{\includegraphics{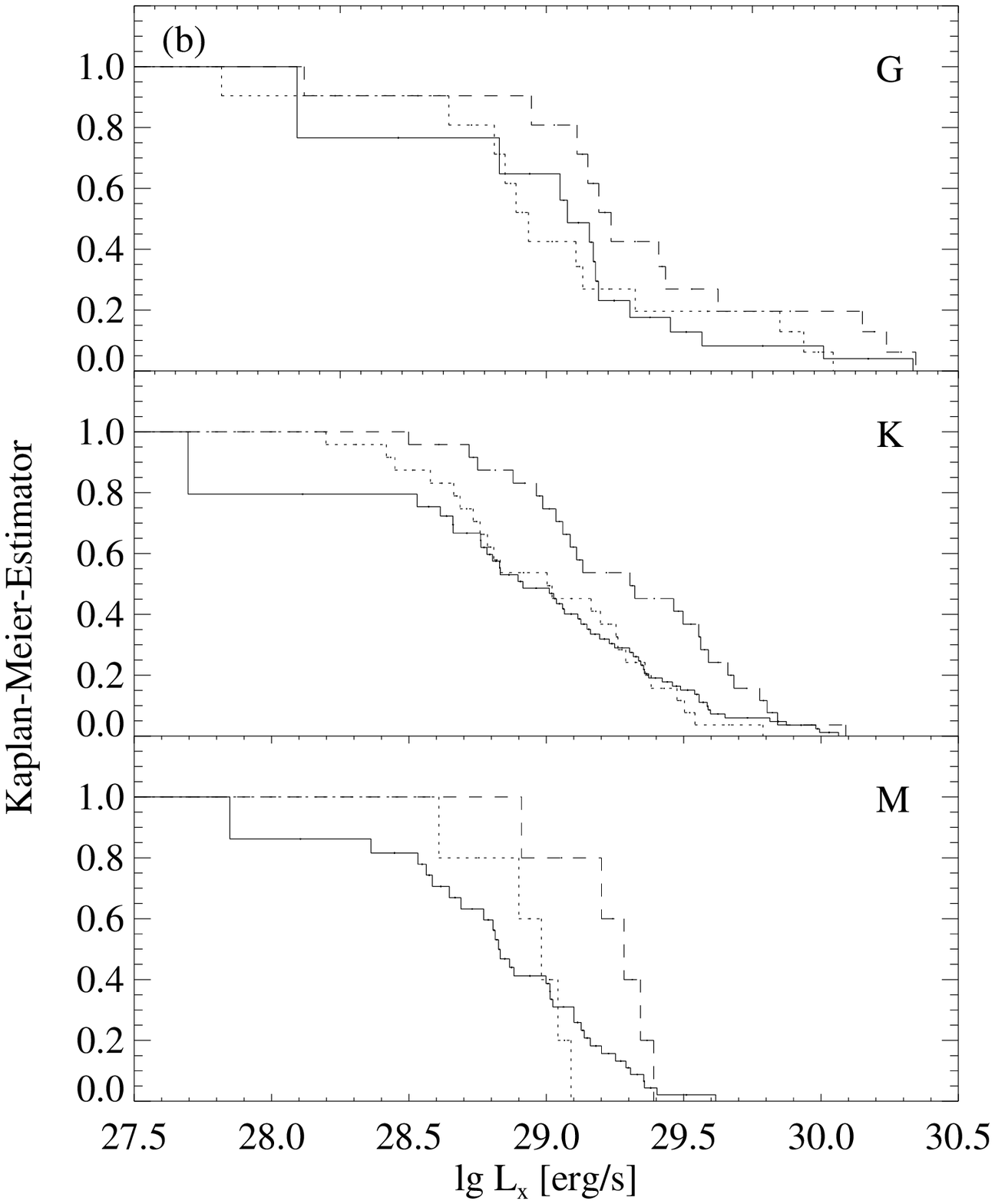}}}
\parbox{9cm}{\resizebox{9cm}{!}{\includegraphics{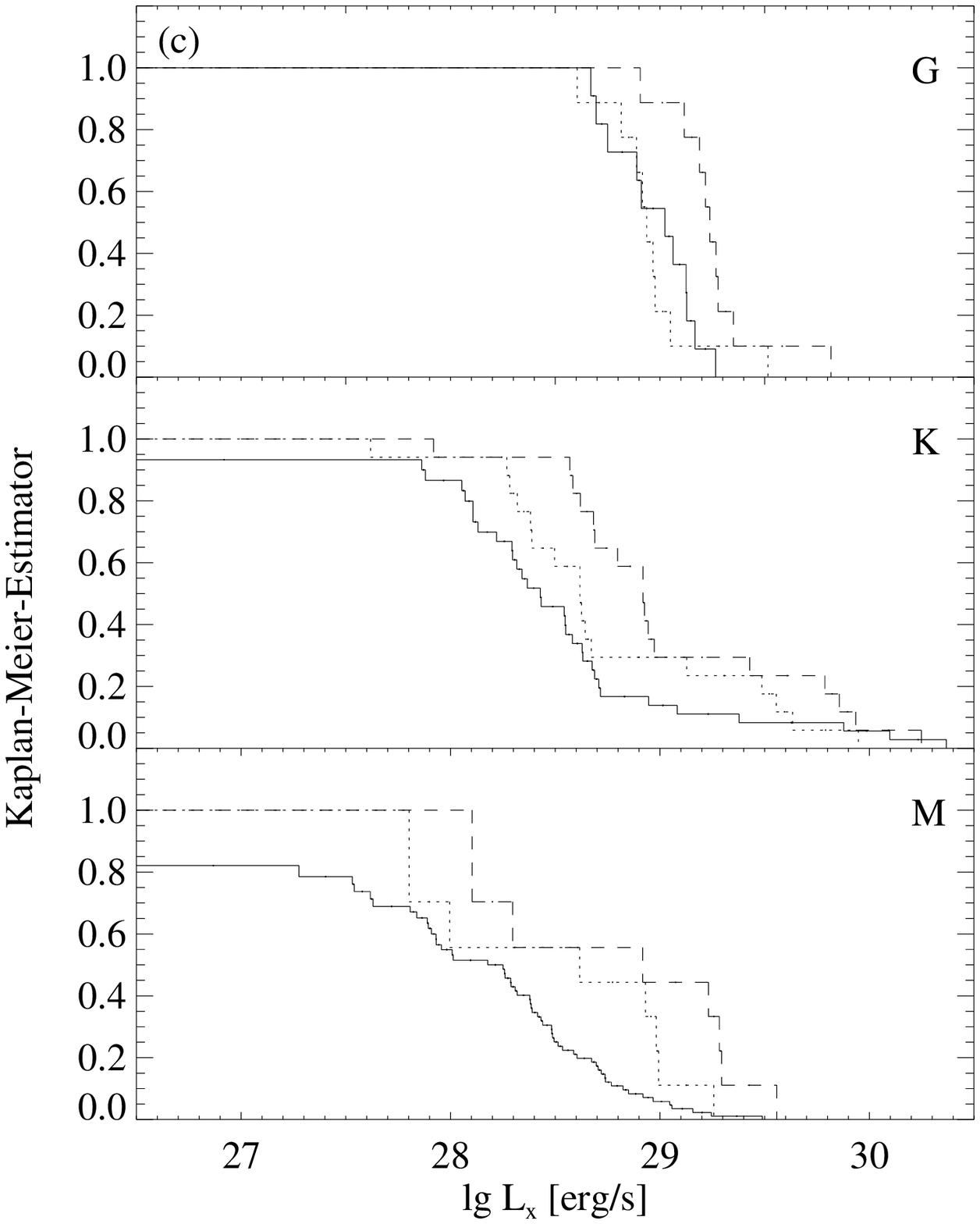}}}
}
\end{center}
\end{figure*}

All XLF presented above may rely to some degree 
on our assumption that all components in 
multiple systems emit X-rays (at the same level). In order to check 
this hypothesis we have studied the XLF of single and binary stars
separately. 
Again we have constructed separate XLF for G, K, and M type stars. In 
Fig.~\ref{fig:ldf_sing_bin}
we show these XLF for TTS, Pleiades and Hyades stars.
For comparison we display also the XLF for binaries derived 
without taking account
of the binary character, i.e. XLF in which each binary has been regarded
as a single star with the observed X-ray luminosity 
(dashed in Fig.~\ref{fig:ldf_sing_bin}). 
Henceforth, these distributions are termed `b1', in contrast to the
distributions `b2' for which equal partition 
of $L_{\rm x}$ onto the components was assumed 
(dotted in Fig.~\ref{fig:ldf_sing_bin}).
As before, binary components with unknown spectral type are not considered.

\begin{table}
\caption{Results of two-sample tests performed with ASURV to distinguish
between the XLF of single and binary stars. For each group (TTS, Pleiads,
and Hyads) and each spectral type we have compared three distributions:
{\em s} - single stars, {\em b1} - binary stars with only one X-ray emitter,
{\em b2} - binary stars assuming that both components emit equal amounts of
X-rays. The probabilities given are for the null-hypothesis that the
compared pair of XLF is drawn from the same parent distribution. We have
applied Gehan's generalized Wilcoxon test (GW), the logrank test, and the
Peto \& Prentice generalized Wilcoxon test.}
\label{tab:ldf_sing_bin}
\begin{center}
\begin{tabular}{lcccc} \hline
Sample & size (ul.) & Prob & Prob & Prob \\
& & GW HV & log rank & P \& Pren. \\ \hline
\multicolumn{5}{c}{\bf TTS K stars} \\ \hline
s - b2  & 34 (11) - 17 (3) & 0.948 & 0.852 & 0.948 \\
s - b1  & 34 (11) - 17 (3) & 0.073 & 0.165 & 0.084 \\ \hline
\multicolumn{5}{c}{\bf TTS M stars} \\ \hline
s - b2  & 60 (28) - 29 (10) & 0.238 & 0.471 & 0.275 \\
s - b1  & 60 (28) - 29 (10) & 0.006 & 0.051 & 0.010 \\ \hline
\multicolumn{5}{c}{\bf Pleiads G stars} \\ \hline
s - b2  & 25 (13) - 16 (5) & 0.844 & 0.953 & 0.789 \\
s - b1  & 25 (13) - 16 (5) & 0.085 & 0.103 & 0.089 \\ \hline
\multicolumn{5}{c}{\bf Pleiads K stars} \\ \hline
s - b2  & 84 (38) - 27 (3) &  0.825 & 0.286 & 0.688 \\
s - b1  & 84 (38) - 27 (3) & 0.002 & 0.001 & 0.004 \\ \hline
\multicolumn{5}{c}{\bf Pleiads M stars} \\ \hline
s - b2  & 60 (29) - 5 (0) & 0.710 & 0.294 & 0.665 \\
s - b1  & 60 (29) - 5 (0) & 0.002 & 0.001 & 0.009 \\ \hline
\multicolumn{5}{c}{\bf Hyads G stars} \\ \hline
s - b2  & 12 (1) - 10 (1) & 0.657 & 0.711 & 0.620 \\
s - b1  & 12 (1) - 10 (1) & 0.003 & 0.005 & 0.005 \\ \hline
\multicolumn{5}{c}{\bf Hyads K stars} \\ \hline
s - b2  & 36 (5) - 18 (1) & 0.134 & 0.095 & 0.150 \\
s - b1  & 36 (5) - 18 (1) & 0.000 & 0.000 & 0.001 \\ \hline
\multicolumn{5}{c}{\bf Hyads M stars} \\ \hline
s - b2  & 89 (35) - 9 (3) & 0.059 & 0.217 & 0.083 \\
s - b1  & 89 (35) - 9 (3) & 0.002 & 0.022 & 0.005 \\ \hline
\end{tabular}
\end{center}
\end{table}
The mean and median of $\lg{L_{\rm x}}$ for all compiled distributions 
are listed in Table~\ref{tab:lxmean}.
Obviously, throughout all
examined groups of stars the distributions `b1' are shifted towards higher
luminosities with respect to the distributions `b2'.
We have performed two-sample tests within ASURV to quantify the
differences. The results are summarized in Table~\ref{tab:ldf_sing_bin}.
The comparison between `s' and `b2' shows in most cases a high probability 
that the distributions are similar. Only for the Hyades K and M stars 
the probability that the distributions of singles and `b2' are different
is $\sim 90$\%.
All samples `s` and `b1', on the contrary, have high probability for 
different underlying parent distributions.

The XLF of Hyades stars have first been examined by \citey{Pye94.1} on the
basis of {\em ROSAT} observations. Their finding that Hyades dK binaries
are X-ray brighter than single Hyads of the same spectral type were
confirmed by \citey{Stern95.1} on a larger sample. Our analysis shows
that the comparison depends sensitively on the way in which 
binary stars are treated. The effect is strongly reduced if it
is assumed that both components in binaries emit X-rays (`b2') with
respect to distributions of type `b1' examined by \citey{Pye94.1} 
and \citey{Stern95.1}. 

\section{Rotation-Activity Relations}\label{sect:x_rot}

The rotation-activity connection has been extensively studied by 
\citey{Walter81.1}, \citey{Walter81.2}, \citey{Walter82.1}, 
\citey{Bouvier90.1}, \citey{Damiani91.1}, \citey{Grankin93.1}, N95, 
\citey{Bouvier97.2}, and \citey{Wichmann98.1}.

In this section we study the subsample of the stars from 
Tables~2~to~7 with measured rotation periods
$P_{\rm rot}$ or projected rotational velocity \vsini.
The choice of the best parameters describing the activity-rotation
relation is not undisputed. We have, therefore, examined different parameter
combinations. On the X-ray side we use the luminosity $L_{\rm x}$, the
surface flux $F_{\rm s}$, and the ratio between X-ray and bolometric
luminosity $L_{\rm x}/L_{\rm bol}$ to characterize the stars. Each star is represented
by its mean X-ray luminosity or upper limit to $L_{\rm x}$ as described in 
Sect.~\ref{sect:ldfs}.
For binaries only one component is considered, because spectral types and
rotation rates are in most cases known only for the primary.
The stellar radii used to compute $F_{\rm s}$
were determined from the Stefan-Boltzmann law.
The rotation is described by the projected rotational velocity, \vsini ,
or the rotation period, $P_{\rm rot}$. 
$P_{\rm rot}$ and \vsini of Pleiades and Hyades stars are listed in the
Open Cluster Data Base. Values for the rotation rates of TTS are taken from
N95, \citey{Bouvier97.2}, and \citey{Wichmann98.1}. 

\begin{table*}
\begin{center}
\caption{Results of statistical tests with ASURV 
for the relation between X-ray emission
and stellar rotation for TTS, Pleiads, and Hyads. The first two columns
are the names of the two parameters to be compared. Next is the size of the
sample, $N$, 
and in brackets the number of upper limits, $N_{\rm lim}$, 
to the rotation and X-ray
parameter. Columns~5~and~6 give the probability that there is {\em no}
correlation between the two parameters according 
to Kendall's and Spearman's test. The slope of a linear regression to the
pair of parameters is given in column~7. For doubly censored data we have
used the linear regression method of \protect\citey{Schmitt85.1}. All
samples where $P_{\rm rot}$ is the rotation parameter have upper limits
only in the X-ray parameters, and the EM algorithm is used.}
\label{tab:corr_x_rot}
\begin{tabular}{llrrrrr}\hline
Par 1 & Par 2 & $N$ & $N_{\rm lim}$ & Kendall & Spearman &
\multicolumn{1}{c}{slope} \\ \hline
\multicolumn{7}{c}{\bf TTS} \\ \hline
\lgvsini & $\lg{L_{\rm x}}$                         & 65  & (0/17) & 0.0031 & 0.0047 & $1.08 \pm 0.36$ \\
\lgvsini & $\lg{F_{\rm s}}$                         & 52  & (0/14) & 0.0009 & 0.0011 & $1.57 \pm 0.46$ \\
\lgvsini & $\lg{(L_{\rm x}/L_{\rm bol})}$           & 52  & (0/14) & 0.0053 & 0.0040 & $1.22 \pm 0.44$ \\ 
$\lg{P_{\rm rot}}$ & $\lg{L_{\rm x}}$               & 39  & (0/7)  & 0.0000 & 0.0001 & $-1.52 \pm 0.39$ \\
$\lg{P_{\rm rot}}$ & $\lg{F_{\rm s}}$               & 38  & (0/6)  & 0.0000 & 0.0000 & $-1.93 \pm 0.39$ \\
$\lg{P_{\rm rot}}$ & $\lg{(L_{\rm x}/L_{\rm bol})}$ & 38  & (0/6)  & 0.0001 & 0.0002 & $-1.49 \pm  0.42$ \\ \hline
\multicolumn{7}{c}{\bf Pleiades} \\ \hline
\lgvsini & $\lg{L_{\rm x}}$                         & 164 & (6/53) & 0.0000 & 0.0000 & $0.61$ \\
\lgvsini & $\lg{F_{\rm s}}$                         & 164 & (6/53) & 0.0000 & 0.0000 & $0.67$ \\
\lgvsini & $\lg{(L_{\rm x}/L_{\rm bol})}$           & 164 & (6/53) & 0.0000 & 0.0000 & $0.88$ \\ 
$\lg{P_{\rm rot}}$ & $\lg{L_{\rm x}}$               & 46  & (0/13) & 0.0008 & 0.0005 & $-0.42 \pm 0.11$ \\
$\lg{P_{\rm rot}}$ & $\lg{F_{\rm s}}$               & 46  & (0/13) & 0.0000 & 0.0001 & $-0.52 \pm 0.11$ \\
$\lg{P_{\rm rot}}$ & $\lg{(L_{\rm x}/L_{\rm bol})}$ & 46  & (0/13) & 0.0000 & 0.0000 & $-0.66 \pm  0.12$ \\ \hline
\multicolumn{7}{c}{\bf Hyades} \\ \hline
\lgvsini & $\lg{L_{\rm x}}$                         & 67  & (41/2) & 0.0008 & 0.0000 & $1.58$ \\
\lgvsini & $\lg{F_{\rm s}}$                         & 67  & (41/2) & 0.0000 & 0.0001 & $1.56$ \\
\lgvsini & $\lg{(L_{\rm x}/L_{\rm bol})}$           & 67  & (41/2) & 0.0003 & 0.0089 & $1.64$ \\ 
$\lg{P_{\rm rot}}$ & $\lg{L_{\rm x}}$               & 21  & (0/2)  & 0.0003 & 0.0004 & $-1.13 \pm 0.32$ \\
$\lg{P_{\rm rot}}$ & $\lg{F_{\rm s}}$               & 21  & (0/2)  & 0.0016 & 0.0016 & $-0.94 \pm 0.28$ \\
$\lg{P_{\rm rot}}$ & $\lg{(L_{\rm x}/L_{\rm bol})}$ & 21  & (0/2)  & 0.3515 & 0.3489 & $-1.51 \pm  0.32$ \\ \hline
\end{tabular}
\end{center}
\end{table*}
For the statistical analysis 
cTTS and wTTS have been combined to yield a larger sample, although
generally wTTS are faster rotators
than cTTS, and they are more X-ray luminous. 
A linear regression has been fitted to all pairs of rotation-activity
combinations using the ASURV EM algorithm or the method by \citey{Schmitt85.1}
for doubly censored data.
In Table~\ref{tab:corr_x_rot} we summarize the results of all correlation
tests, and also give the slopes of the linear regression.
According to the statistical tests 
X-ray emission and rotation are clearly correlated for most of the examined
stellar samples.
For a given X-ray parameter the probability for a correlation with 
$P_{\rm rot}$ is in most cases 
larger than the probability for a correlation with \vsini. 
This is probably due to the unknown inclination angle in \vsini 
whose arbitrary orientation tends to destroy
any intrinsic correlation between the rotation and X-ray emission. Using 
$P_{\rm rot}$ should therefore be
more meaningful. However, measurements of the actual periods (by spot 
modulation) are much sparser than spectroscopic observations of \vsini,
leading to a smaller data set.

\begin{figure}
\begin{center}
\resizebox{9cm}{!}{\includegraphics{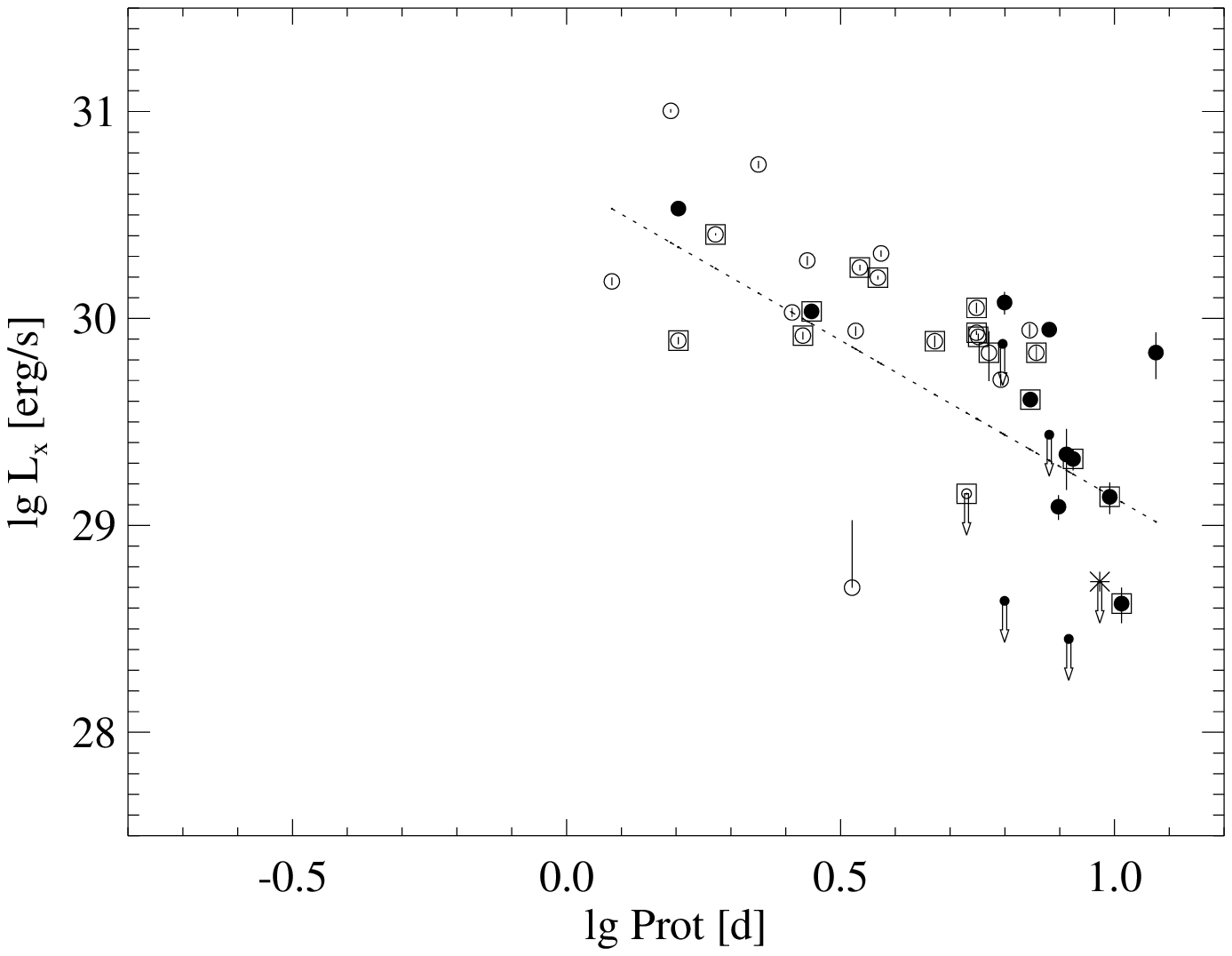}}
\resizebox{9cm}{!}{\includegraphics{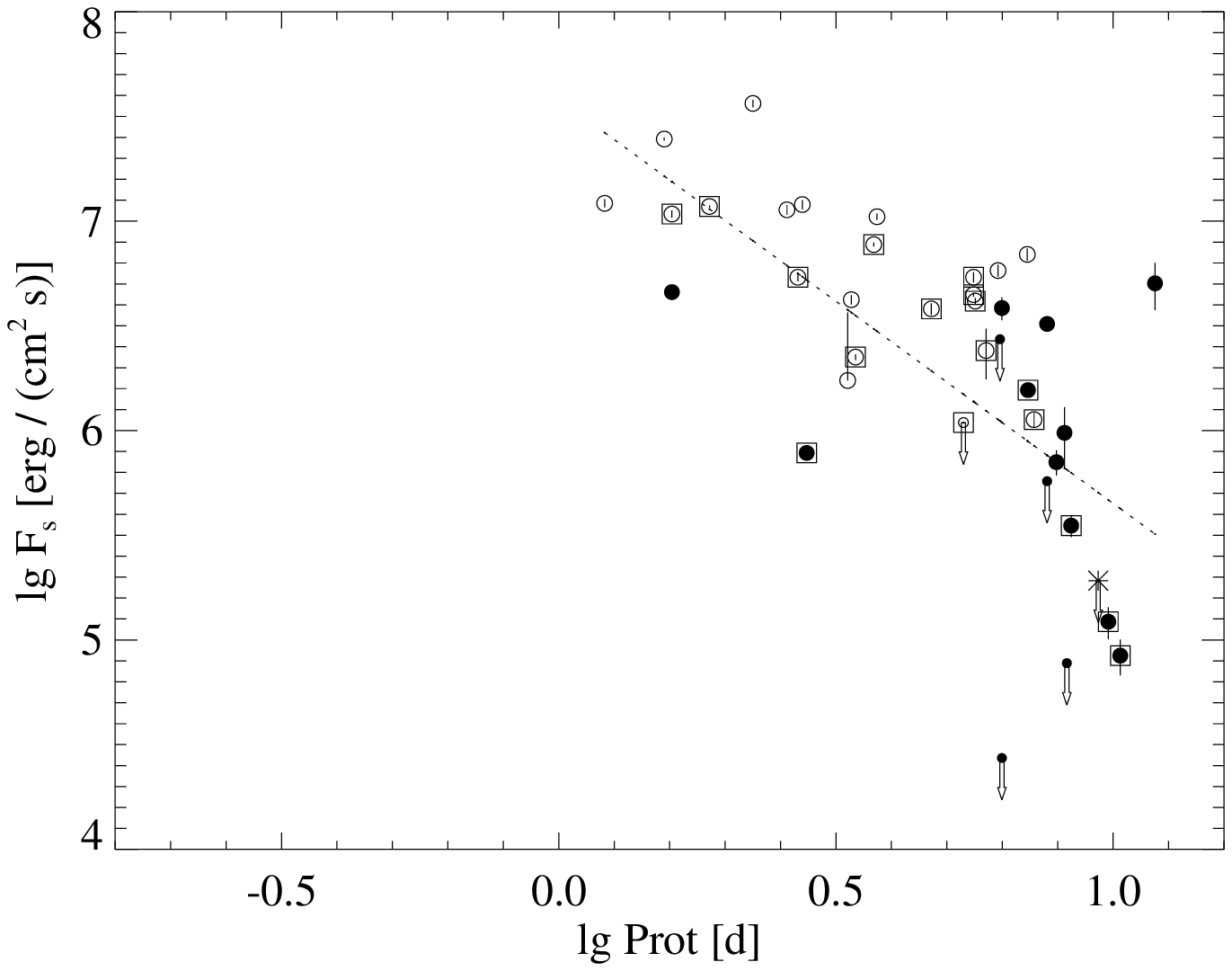}}
\resizebox{9cm}{!}{\includegraphics{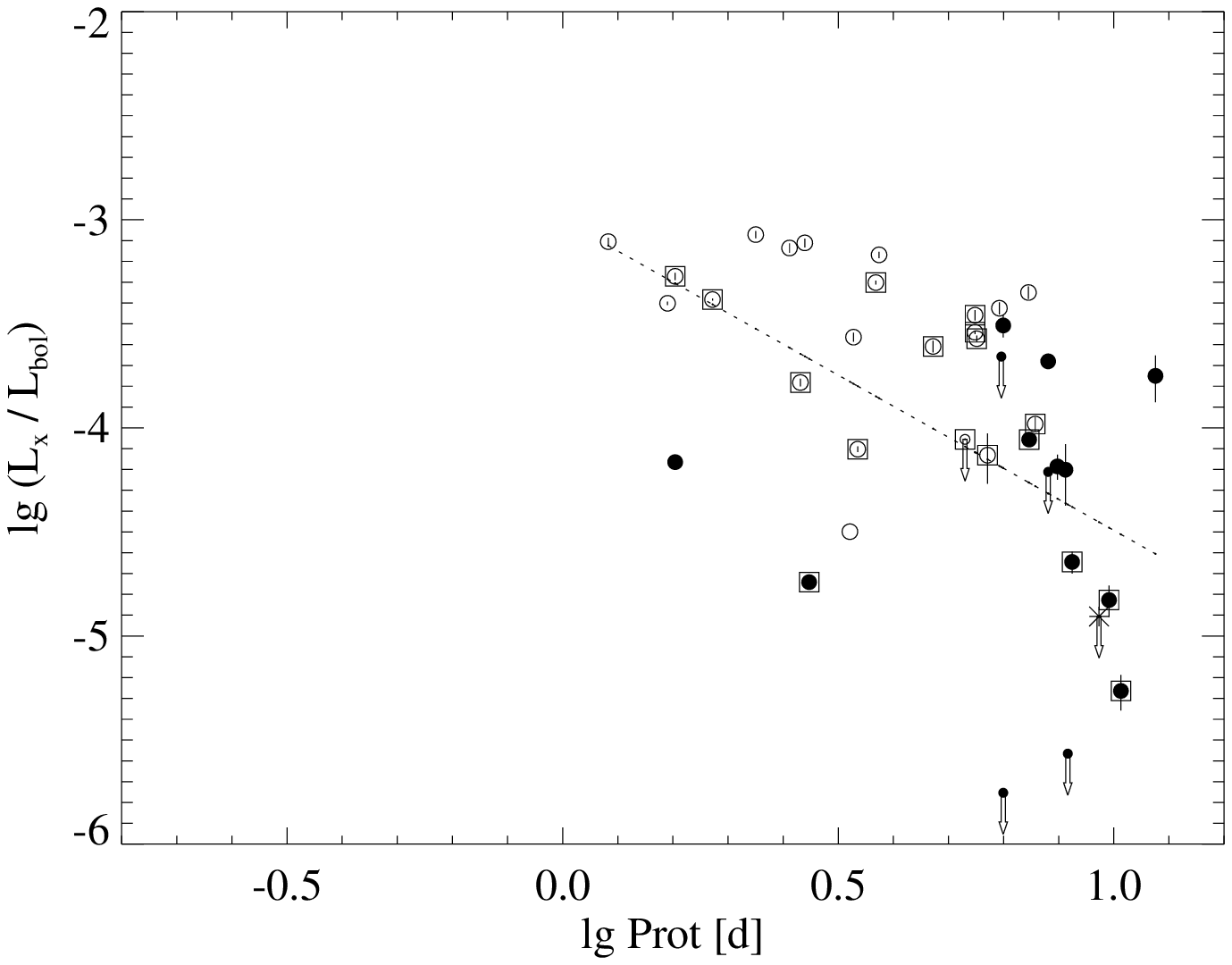}}
\caption{Relation between the rotation period and
different X-ray parameters for TTS from the Taurus-Auriga region: {\em
top} -- X-ray luminosity, {\em middle} -- X-ray surface flux, and {\em
bottom} -- Ratio between X-ray luminosity and bolometric luminosity. cTTS are
represented by filled symbols and wTTS by open symbols. TTS of unknown
nature are displayed as asterisks. Multiple stars are marked by boxes. The
solid lines are linear regressions computed with the EM algorithm
implemented in ASURV. The size of the errors bars varies a lot due to very
different {\em ROSAT} exposure times, and they are sometimes smaller than
the plotting symbol.}
\label{fig:x_prot_tts}
\end{center}
\end{figure}

\begin{figure}
\begin{center}
\resizebox{9cm}{!}{\includegraphics{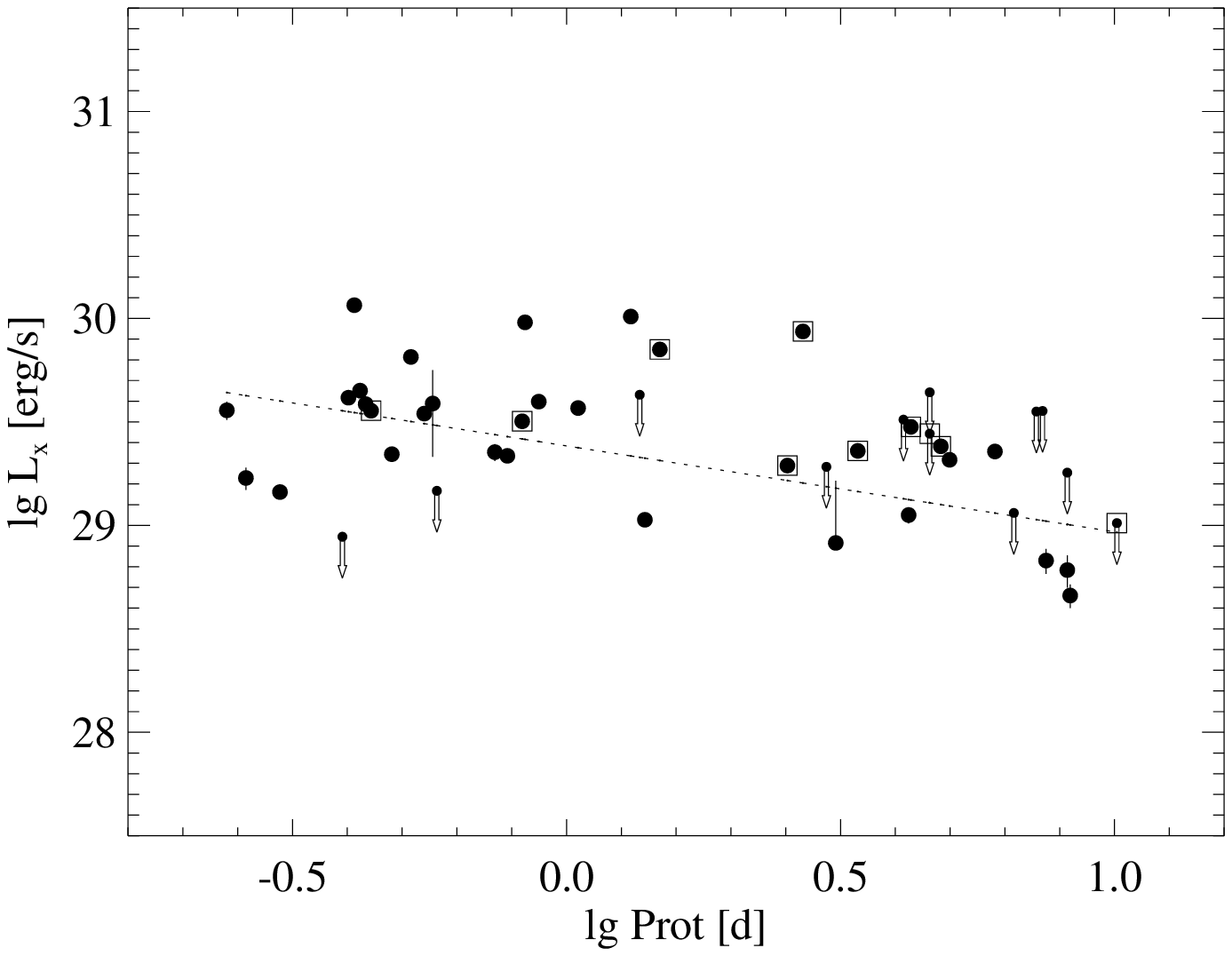}}
\resizebox{9cm}{!}{\includegraphics{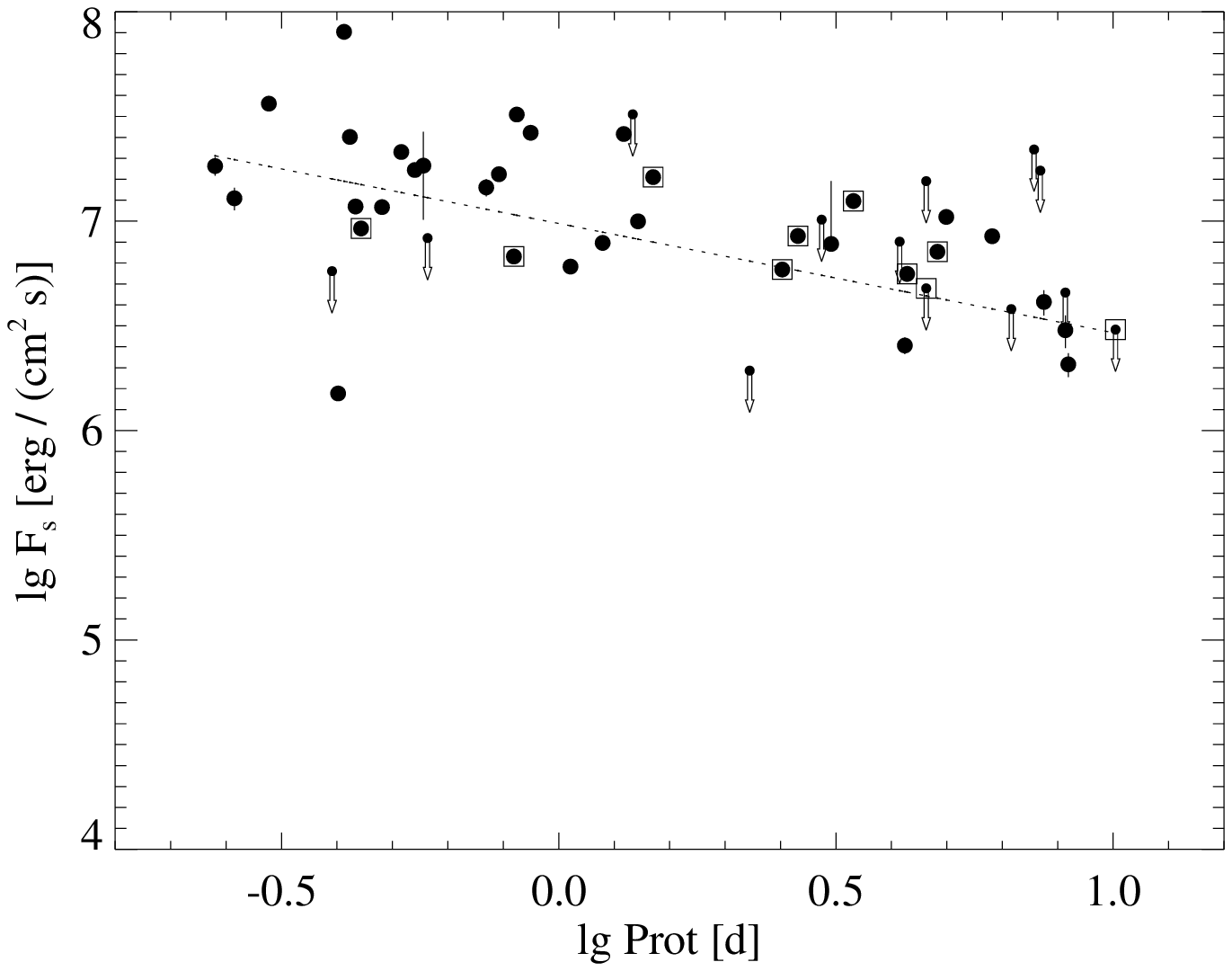}}
\resizebox{9cm}{!}{\includegraphics{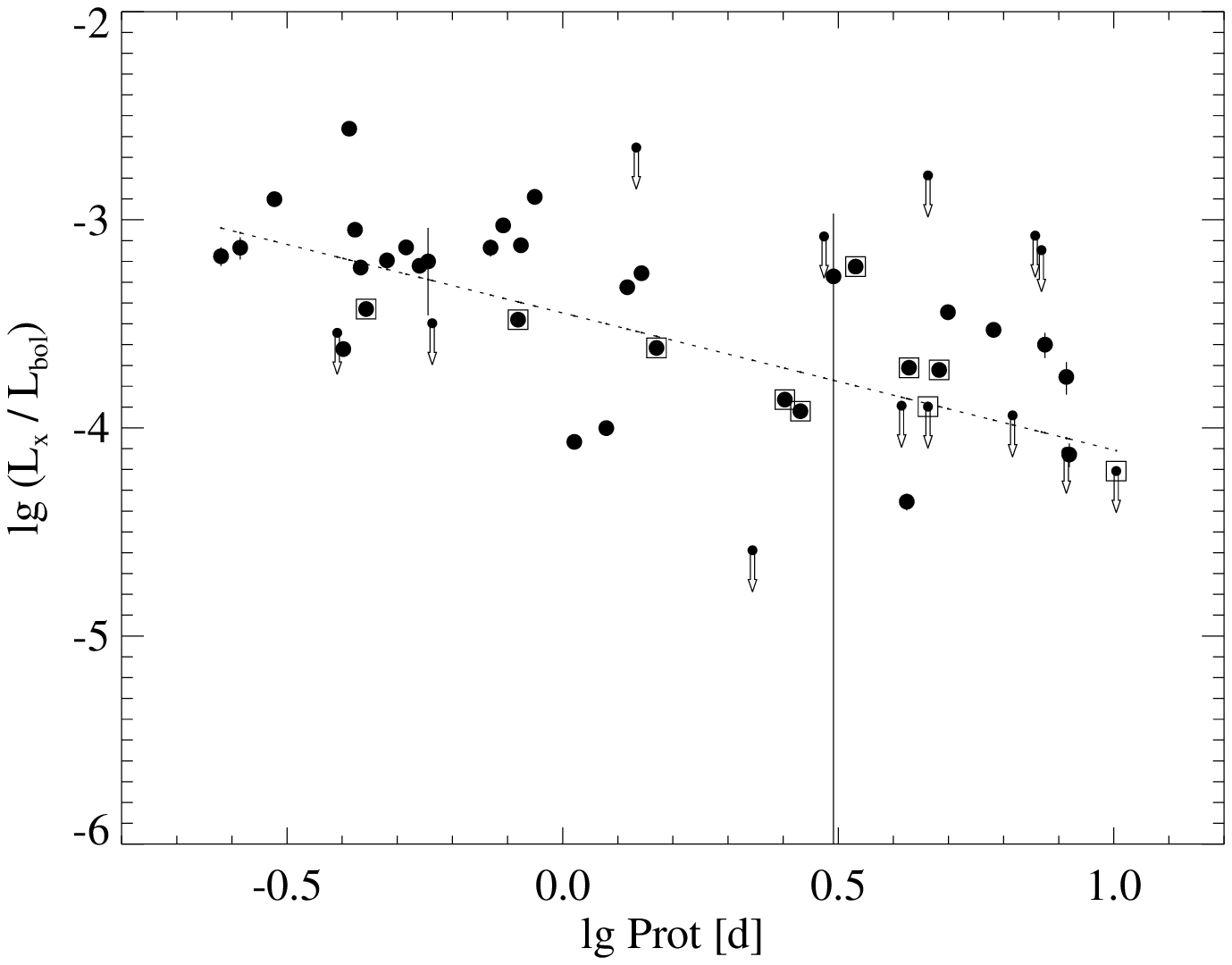}}	
\caption{Same as Fig.~\ref{fig:x_prot_tts} for the Pleiades.}
\label{fig:x_prot_ple}
\end{center}
\end{figure}

\begin{figure}
\begin{center}
\resizebox{9cm}{!}{\includegraphics{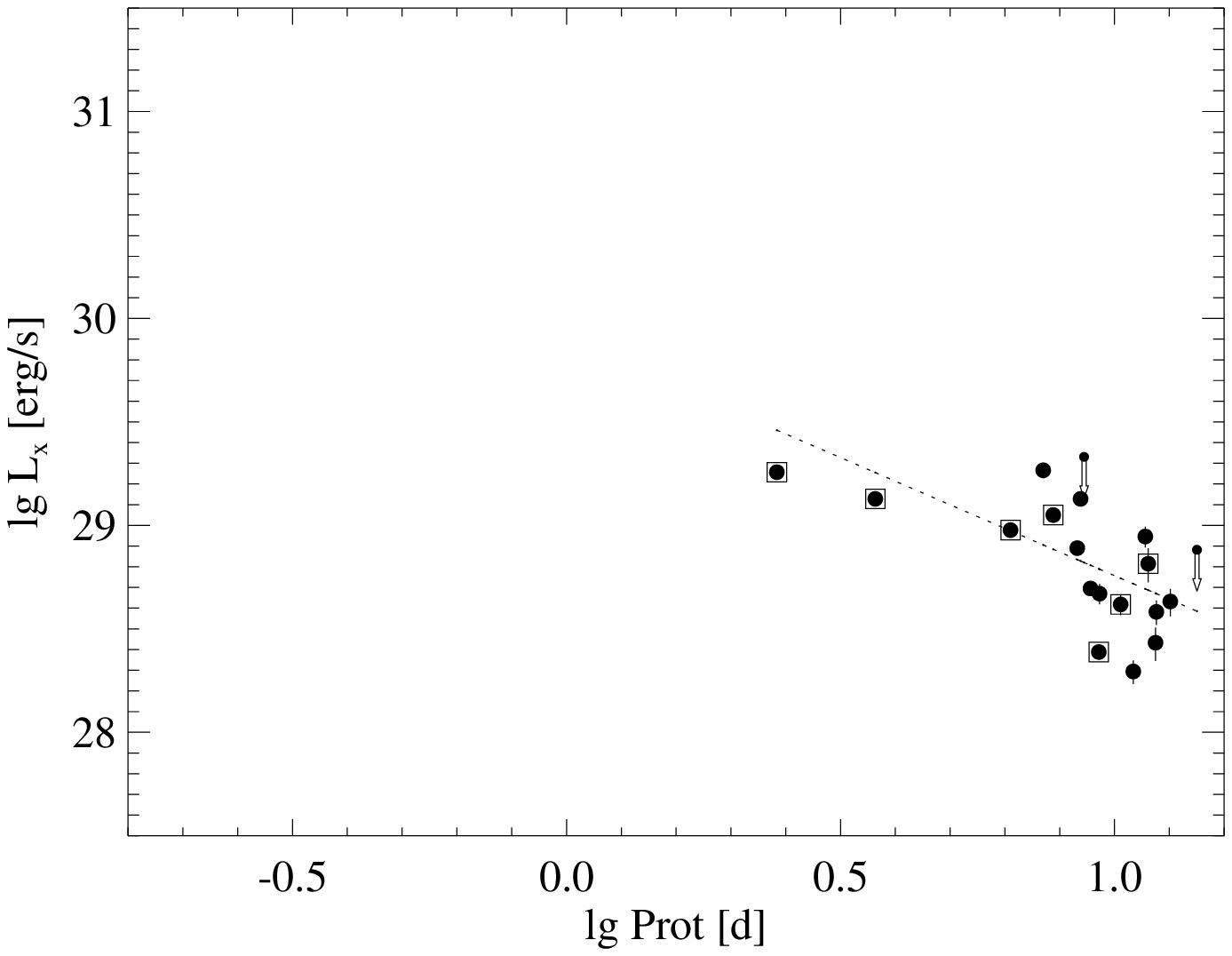}}
\resizebox{9cm}{!}{\includegraphics{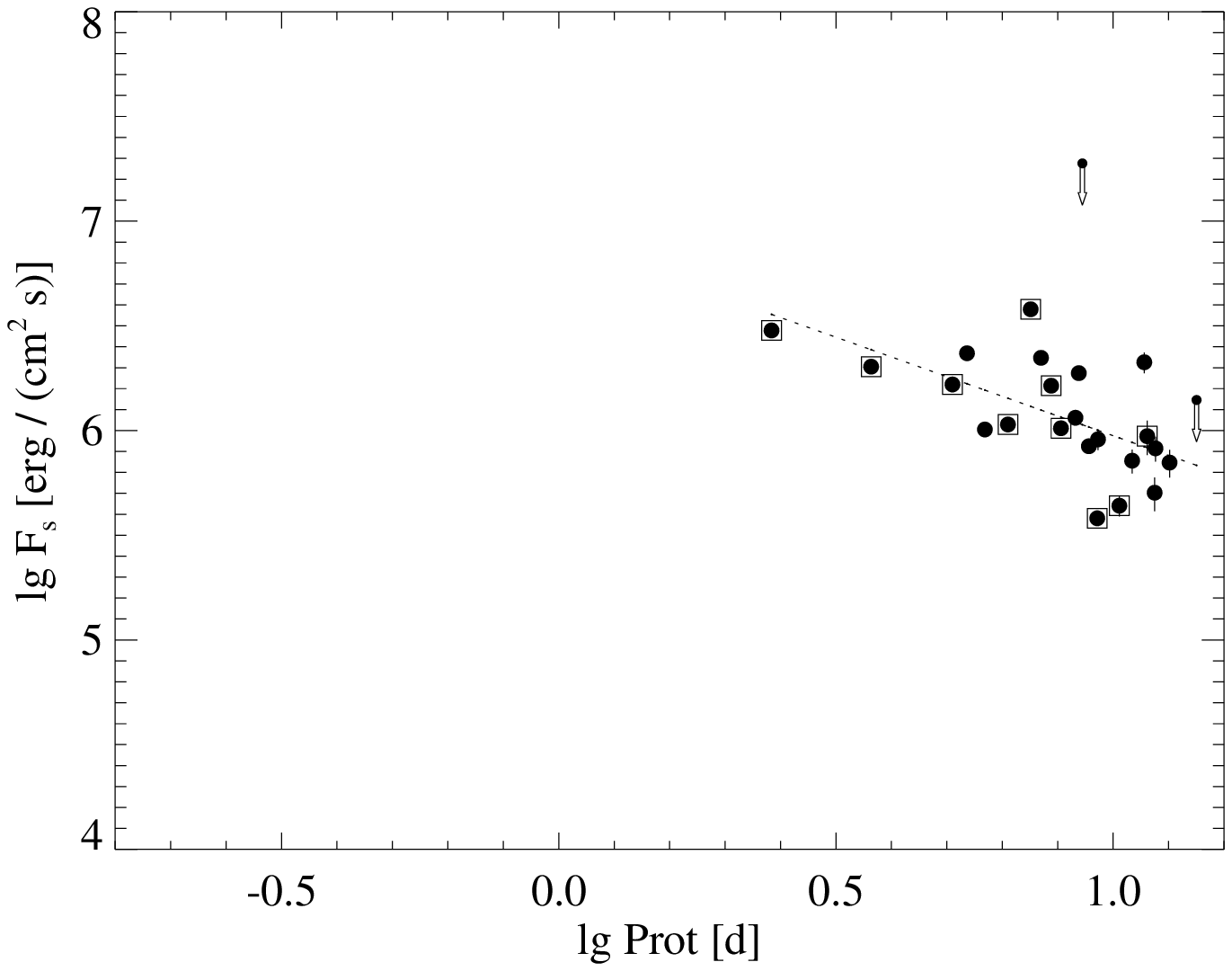}}
\resizebox{9cm}{!}{\includegraphics{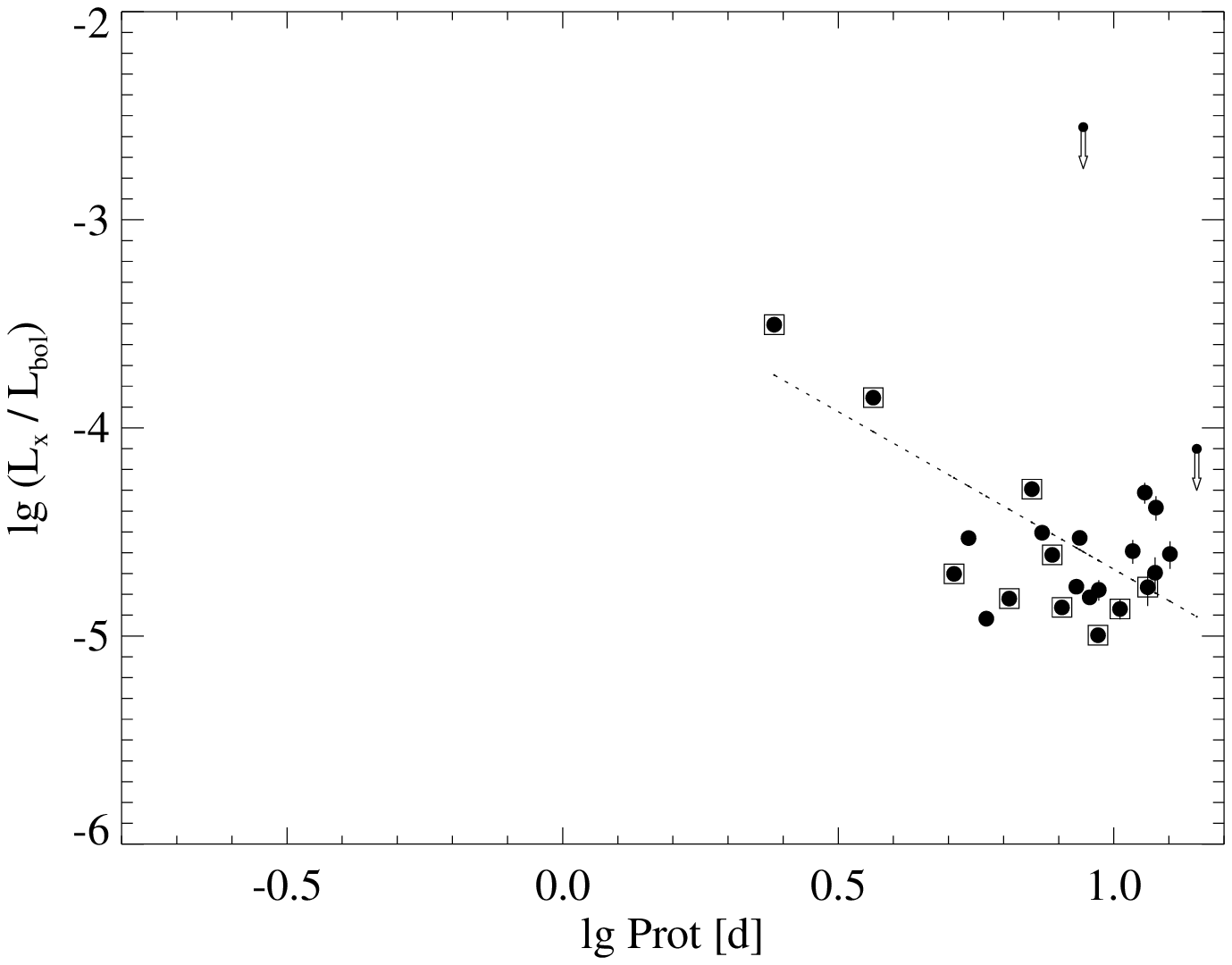}}	
\caption{Same as Fig.~\ref{fig:x_prot_tts} for the Hyades.}
\label{fig:x_prot_hya}
\end{center}
\end{figure}

We show correlations of all possible combinations of the above mentioned 
X-ray parameters with $P_{\rm rot}$ in 
Figs.~\ref{fig:x_prot_tts}~to~\ref{fig:x_prot_hya} for TTS, Pleiads, and
Hyads. Overlaid are the linear regressions corresponding to the  
power law relation from Table~\ref{tab:corr_x_rot}.
The lowest significance is found in the Hyades. This may however be due
to the limited range in rotation period (only two stars with $P_{\rm rot} <
4\,{\rm d}$), and because the
Hyades with known period have a small range of spectral types.

\section{Discussion}\label{sect:discussion}

\subsection{The XLF of cTTS and wTTS}\label{subsect:disc_xldf}

We have reanalysed the XLF for cTTS and wTTS
in Taurus-Auriga, first presented by N95, increasing the sensitivity
with respect to the RASS by $\sim 2$ orders of magnitude.
Our pointed PSPC observations confirm that in Taurus-Auriga wTTS 
are on average more X-ray luminous than cTTS. This is in contrast 
to studies of Cha I and $\rho$ Oph
(\cite{Feigelson93.1}, \cite{Casanova95.1}, \cite{Grosso00.1}), 
where no difference was
found between the two sub-classes of TTS concerning their X-ray emission
level. In a study of the Orion Nebula region with the {\em ROSAT} HRI
\citey{Gagne95.1} found slightly lower median $L_{\rm x}$ and $L_{\rm
x}/L_{\rm bol}$ values for stars with massive accretion disks,
i.e. cTTS. 
\citey{Alcala97.1} have found higher X-ray luminosities for
{\em ROSAT} discovered wTTS in the outer parts of the Cha I and Cha II regions.
This seems to indicate that samples of wTTS may be biased towards strong
X-ray emitters, and that discrepancies can arise from the different spatial
distribution of the cTTS and wTTS sample.

We have ruled out such an X-ray selection bias for our sample, by comparing
the XLF for wTTS discovered by means of their X-ray emission to those
which have been identified in other ways. 
XLF constructed for a coeval subgroup of cTTS and wTTS located
in a central portion of the Taurus-Auriga complex, the L1495E cloud,
show the same disagreement. Therefore, the difference 
does not seem to be related to the wide spatial extension (hence large
age spread) 
of the Taurus star forming region. In addition this test shows that the
disagreement is not caused by the different sensitivities (due to different
exposure times) of the various combined PSPC observations. 

Further effects, like different spectral type distribution, the
specific choice of the $W_{\rm H\alpha}$ boundary between cTTS and wTTS, or
our way of splitting the X-ray emission on all components in multiples, can not
explain the observed discrepancies between the cTTS and wTTS XLF.
To investigate
whether the high number of upper limits in the cTTS sample affects the
shape of the XLF we have also computed XLF neglecting all upper limits. 
(\citey{Grosso00.1} have not included 
upper limits in their XLF of $\rho$ Oph.) 
The structure of the XLF, however, remains unaffected.

We conclude that there is an intrinsic
difference in X-ray emission from cTTS and wTTS in Taurus.
Besides the extinction effect discussed above the different evolutionary
state of TTS in different star forming regions may contribute to the
observed discrepancies.
It should be noted that the subsamples of cTTS and wTTS in Taurus with known 
$T_{\rm eff}$ and $L_{\rm bol}$ occupy the same region
in the H-R diagram, 
i.e. the difference in $L_{\rm x}$ seems not to be a direct 
age effect. 

The correlation between the X-ray luminosity and $P_{\rm rot}$ 
we found for all examined samples 
suggests that the X-ray emission level may be governed by rotation.
To check this hypothesis we have computed separate XLF for fast rotating
wTTS ($\vsini > 22\,{\rm km/s}$, the mean \vsini for wTTS), and slowly
rotating wTTS ($\vsini < 12\,{\rm km/s}$). Indeed, the slow rotators are
characterized by lower X-ray luminosity 
($\lg{L_{\rm x,mean}} = 29.54 \pm 0.13$
versus $30.00 \pm 0.11$ for the fast group). This explains some but not all
of the discrepancy between the XLF of Fig.~\ref{fig:ldf_rass_point}. 
From the mean
rotation rate of cTTS and wTTS and the mean \lgLx values derived from the
KME analysis the slope in Fig.~\ref{fig:x_prot_tts} would be expected to be much
steeper. But note, that only a small fraction of TTS has measured
rotation periods, and the large spread in the observed rotation-activity
relation may be due to mixing of stars with different mass. 

If, indeed, rotation is the major parameter that determines the amount
of X-rays emitted by a given star then cTTS and wTTS in Taurus-Auriga
are expected to have different $L_{\rm x}$ 
because the wTTS are on average faster rotators (see \cite{Bouvier93.1} and
our Fig.~\ref{fig:x_prot_tts}). Different distributions of rotation 
periods are also found in other star forming regions, e.g. Lupus 
(\cite{Wichmann98.2}). Only in Orion cTTS and wTTS are found to 
rotate at the same speed (\cite{Stassun99.1}). 
The rotational state of the PMS stars in Cha~I and $\rho$ Oph has not
yet been investigated in detail.
We suspect that most of the wTTS in Taurus-Auriga (including those in L1495E)
have spent a longer time than those in Cha~I and $\rho$ Oph 
since they have dispersed their disks, and  
therefore have had more time to spin up, and consequently 
should drive a more powerful dynamo. This implies that the disk lifetimes
depend on the local condition in the star forming region.
We remark that this hypothesis can only be tested
after more measurements of rotational velocities in these different
regions are available. In a later paper we will compare the XLF in 
different star forming regions in more detail.

\subsection{Spectral Type and Age Dependence of the X-ray Emission}\label{subsect:disc_tph}

We have compared the XLF of TTS in Taurus-Auriga, the Pleiades, and the 
Hyades. Following early studies by the {\em EO} the XLF of Pleiades and Hyades
had been examined with the improved sensitivity of {\em ROSAT} 
(see e.g. \cite{Hodgkin95.1} and \cite{Micela96.1}, \cite{Pye94.1}, 
\cite{Stern95.1}). However, all
studies of X-ray luminosity on these young clusters
were based on smaller data sets than the one presented here.

In lack of the knowledge about individual masses we
take account of the known mass dependence of 
the X-ray luminosity by regarding G, K, and M stars separately. 
For all spectral type groups wTTS 
are found to be the strongest X-ray emitters, 
and the Hyades show the lowest level of X-ray emission.
The difference between $\langle L_{\rm x} \rangle$ of the 
Pleiades and the Hyades is small for G stars where the spread in the
mass distribution is largest, but large for M stars which have more uniform
masses. This suggests that the decline in the
X-ray emission is mostly an age effect.
The XLF of cTTS and the Pleiades intersect each other, 
because the Pleiades are
characterized by a much steeper distribution indicating less spread in 
$L_{\rm x}$. This difference 
may be a result of the uniform distance assumed for all stars in 
a given group (except the Hyades for which individual {\em Hipparcos}
parallaxes were used).
If the extension in the direction along the line-of-sight is comparable
to the observed spatial dispersion, the TTS in Taurus-Auriga should be
subject to a distance spread of $\sim$\,50\,pc.
Consequently the luminosities
of some stars are underestimated while others are overestimated, 
thus leading to a larger spread in $L_{\rm x}$ and a flattening of the
XLF. For the more compact Pleiades region instead 
the assumption of uniform distance may be adequate.

The XLF of Hyades K stars show a substructure appearing as an edge
at $\lg{L_{\rm x}} \sim 28.7$. In order to explain this feature 
we have divided the K star
Hyades into two subgroups of $\lg{L_{\rm x}}$ larger/smaller than $28.7$.
No differences between these two samples 
were found concerning the distribution of effective
temperature, distances, and location on the sky. Only few of the Hyades 
K stars have measured \vsini or rotation period. 
Therefore, the hypothesis that the
high-luminosity tail is composed of the fast rotators can not be tested.
Note, that the edge in the slope is seen in both single and binary stars
(see Fig.~\ref{fig:ldf_sing_bin}), but seems to be
more pronounced for single stars. We suggest, that the effect is due
to as yet undiscovered multiples among the K type Hyades. 

We have extended our investigation of the dependence of the X-ray 
emission on spectral type by direct examination of correlations between
these parameters (see Fig.~\ref{fig:act_bv}).
This investigation reveals differences between TTS, Pleiades,
and Hyades which we suppose are related to the different ages of these groups.
For stars on the MS $T_{\rm eff}$ corresponds to mass, and mass
is related to the depth of the convection zone. The observed anti-correlation
between \lgLrat and \lgTeff from Fig.~\ref{fig:act_bv} 
therefore demonstrates the importance of convection for X-ray activity.
Although there is a tendency of 
\lgLrat being larger for cooler stars, the absolute amount of X-rays
emitted is smaller (see Figs.~\ref{fig:ldf_sptypes}~and~\ref{fig:act_bv}).
In the Pleiades $L_{\rm x}$ does not strongly depend on spectral
type, although \lgLrat decreases with increasing $T_{\rm eff}$.
This is most likely due to the shorter time
the latest type stars in the Pleiades have spent on the MS. Most of the
late K and M type Pleiads did not spin down to 
loose their high initial activity level, yet.
The PMS TTS show no correlation between \lgLrat and \lgTeff. This may be
due to the large age spread in the TTS sample ($10^{5..7}$\,yrs).

The most active stars of all groups are characterized by \lgLrat $\sim -3$, 
the canonical value for late-type stars. 
This behavior is been referred to as `saturation', and has been described 
in the literature; see e.g. \citey{Fleming89.1}, 
\citey{Feigelson93.1}, \citey{Micela96.1}, 
\citey{Randich96.1}, \citey{Stauffer97.1}, \citey{Micela99.1}. 
A common explanation is that all saturated stars have reached
their highest possible level of X-ray activity, e.g. by coverage of
the full surface with active regions. The stellar radius rather than rotation would
then determine the X-ray emission level (see \cite{Fleming89.1}). 
The correlation between $L_{\rm x}$ and spectral type in TTS 
may be understood in terms of such a saturation effect: 
Fig.~\ref{fig:act_bv} suggests that many TTS regardless
of their spectral type have reached the saturation level. However, 
the more luminous the stars, the larger they are,
and the higher the saturation level for $L_{\rm x}$. Therefore, 
for given $L_{\rm bol}$ the X-ray luminosity is limited by a value 
that corresponds to saturation, 
and which is lower for later spectral types.

The dispersion of $\lg{L_{\rm x}}$ for given spectral type 
can be regarded from two points of view: 
(a) all stars of given spectral type show
intrinsically similar amounts of X-ray emission, 
and the spread in $L_{\rm x}$
is caused by variability of individual stars, or (b) 
the dispersion reflects different activity levels of the stars.
Our analysis of the longterm X-ray behavior of these stars 
(to be presented in a subsequent paper; Stelzer et~al. in prep.) 
suggests little variability on long timescales 
making the former hypothesis improbable.
The distribution of $L_{\rm x}$ within stars of homogeneous spectral type
thus more likely reflects the variety of X-ray emission from individual stars.

\subsection{Are Hyades Binaries Overluminous ?}\label{subsect:disc_xlf_hyabin}

\citey{Pye94.1} have examined the XLF of Hyades stars combining 11 {\em
ROSAT} PSPC observations. In their sample they found that Hyades dK
binaries are overluminous in X-rays: all binary dK stars analysed by 
\citey{Pye94.1} were brighter than any of the single dK stars. 
This result was confirmed by \citey{Stern95.1} on a larger sample of
Hyades drawn from the RASS. 

In our analysis of the XLF in the Hyades we have treated binary stars
in two ways: (A) in the same way as singles, i.e. without taking
account of the multiplicity (sample `b1'), and (B) dividing the observed
luminosity by two to account for X-rays from both components (sample `b2').
We find a probability of $\sim$\,10-15\% for the 
null-hypothesis that the distributions of singles (`s') and `b2'
among the Hyades K stars are drawn from the same parent distribution. 
For Hyades M
stars (not examined by \cite{Pye94.1} due to lack of statistics but found
to display a similar though less pronounced divergence 
between single and binary XLF in the study of \cite{Stern95.1}) we find a 
similar probability for the rejection of the null-hypothesis that
`s' and `b2' are drawn from the same parent distribution. 
However, the sample of M star binaries in the Hyades is very small 
(9 stellar systems). 
For all other pairs of `s' - `b2' distributions, i.e. those of Hyades
G stars, Pleiades, and TTS, there is no statistical evidence for 
differences.
The agreement between the XLF of single (`s') and binary (`b2') stars 
is expected if the components
in binaries have no mutual influence on their activity, and if indeed the
distribution of the observed X-ray emission equally on all components 
conforms with the real situation. This seems likely 
because binaries with very high mass ratio, i.e. largely
different $L_{\rm x}$, are more difficult to detect than equal mass ratio 
binaries.

When compared to the distributions `b1', singles are fainter in all cases 
(probability for the distributions being similar $<\,10$\%).
This is in agreement with the study of \citey{Pye94.1} and
\citey{Stern95.1} who have examined samples of type `b1'.

This results emphasize that it is important to consider the binary
character when analysing XLF of double stars. Splitting the X-ray emission
onto the components significantly decreases the difference between 
single and binary XLF. However, some discrepancy for the Hyades K
and M stars remains unexplained. 
A proper treatment of binary stars 
is also important in correlation studies, as it decreases the spread.

\subsection{The Age-Activity-Rotation Connection}\label{subsect:aar_connection}

\begin{figure}
\begin{center}
\resizebox{9cm}{!}{\includegraphics{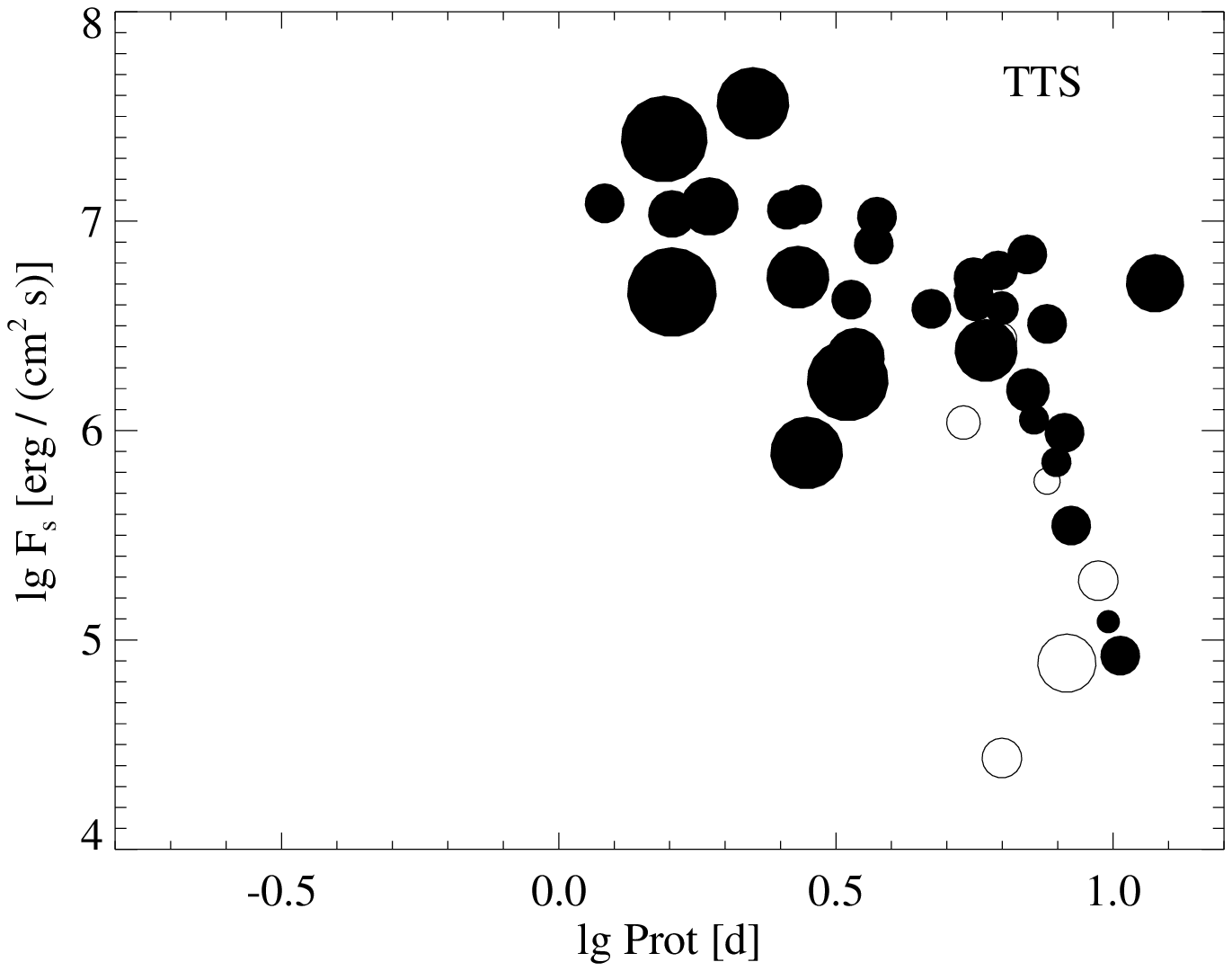}}
\resizebox{9cm}{!}{\includegraphics{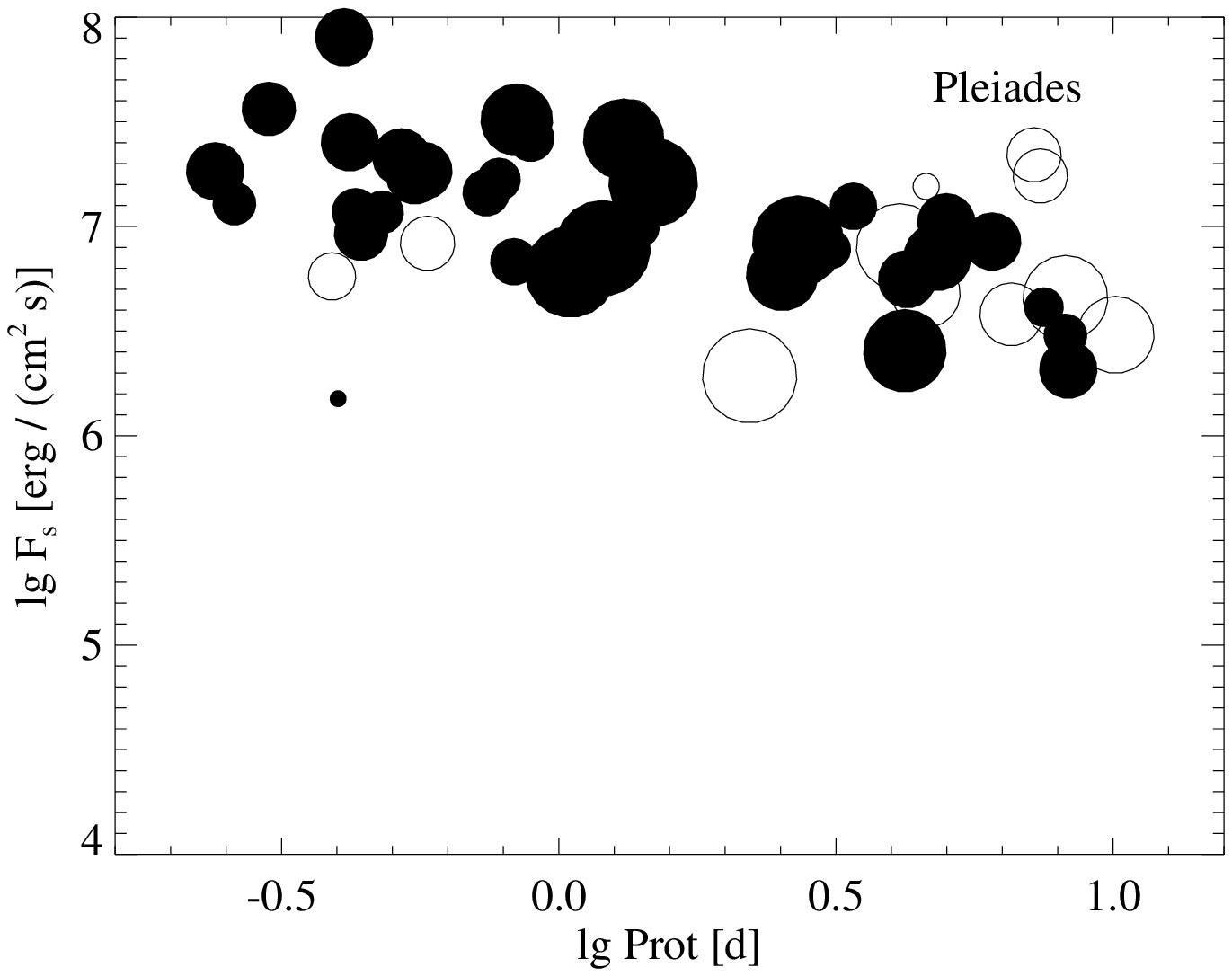}}
\resizebox{9cm}{!}{\includegraphics{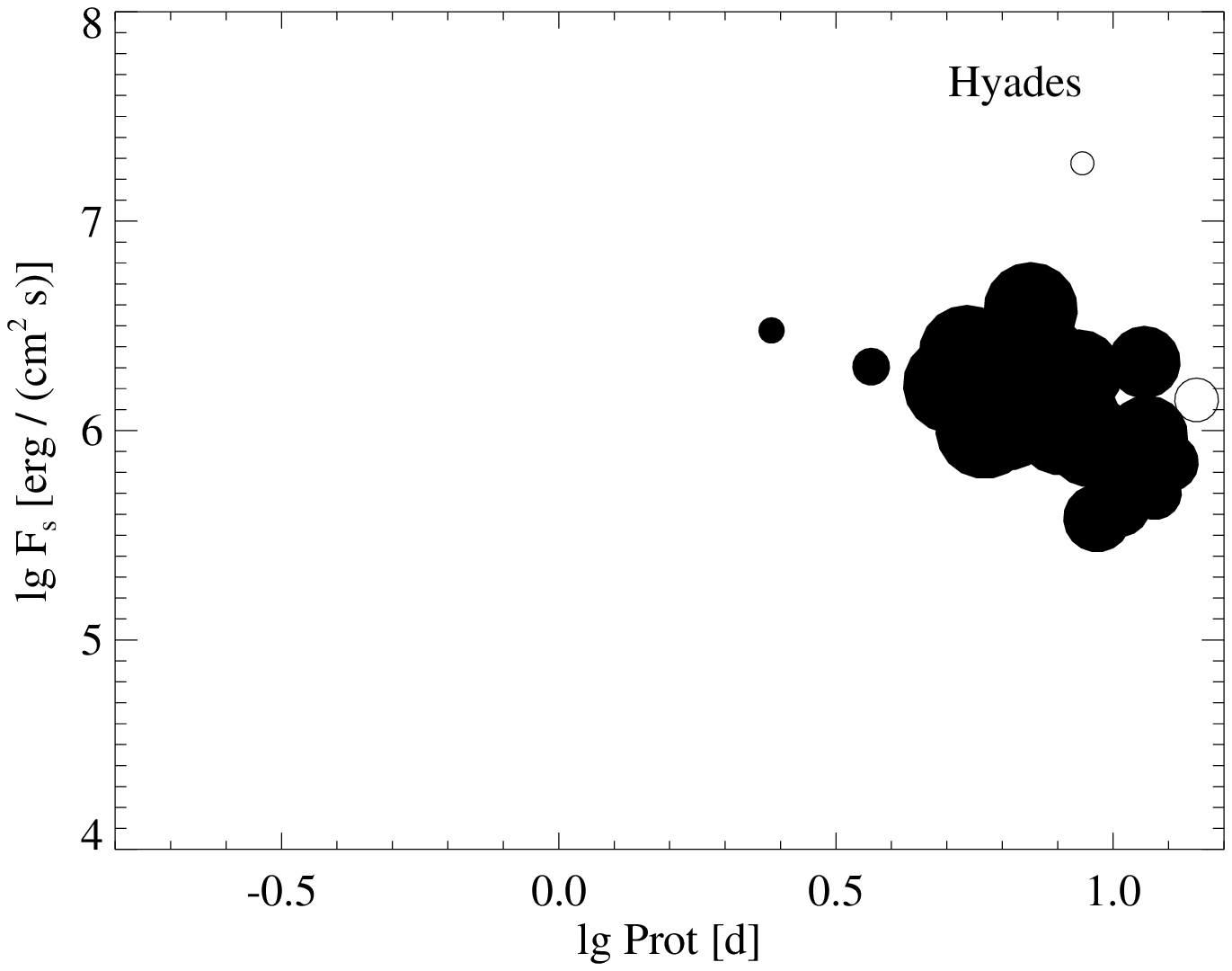}}
\caption{X-ray surface flux versus rotation period for TTS, Pleiades, and 
Hyades indicating the distribution of effective temperatures (plotting
symbols are scaled to $T_{\rm eff}$). Note, that most of the slow rotators
in the TTS sample are cool objects. Open circles are upper limits for
undetected objects.}
\label{fig:fs_prot_teff}
\end{center}
\end{figure}

We have shown that the rotation period and various measures for 
the X-ray activity (i.e. luminosity, surface flux, 
and $L_{\rm x}/L_{\rm bol}$-ratio) are correlated for all examined age
groups. The steepness of the activity-rotation
relation is very different for TTS, Pleiades, and Hyades, with the
largest slope for the TTS, e.g. slow rotators in the 
Pleiades have much higher surface flux
than TTS with similar periods 
(see Figs.~\ref{fig:x_prot_tts}~to~\ref{fig:x_prot_hya}). 
We think that these differences can be
explained by the particular distribution of spectral types: In 
Fig.~\ref{fig:fs_prot_teff} we show the $\lg{F_{\rm s}} - \lg{P_{\rm rot}}$
diagrams with plotting symbols scaled according to $T_{\rm eff}$. In the TTS
sample we observe a clear clustering of cooler stars at slow rotation periods. 
For given $L_{\rm bol}$ and $L_{\rm x}$ 
cooler stars have larger radius and therefore smaller surface flux. This
results in a steeper slope for the TTS sample. 

Fast rotators are found at all spectral types in the Pleiades and among
the TTS. 
Indeed, the fastest rotators form the upper
envelope to the \lgLrat - \lgTeff diagram (Fig.~\ref{fig:act_bv}). 
At the age of the Hyades most stars (regardless of spectral type) have 
slowed down their rotation, such that the range of measured periods is
limited, and definitive statements about 
the activity-rotation connection for the Hyades are difficult.

\begin{figure}
\begin{center}
\resizebox{9cm}{!}{\includegraphics{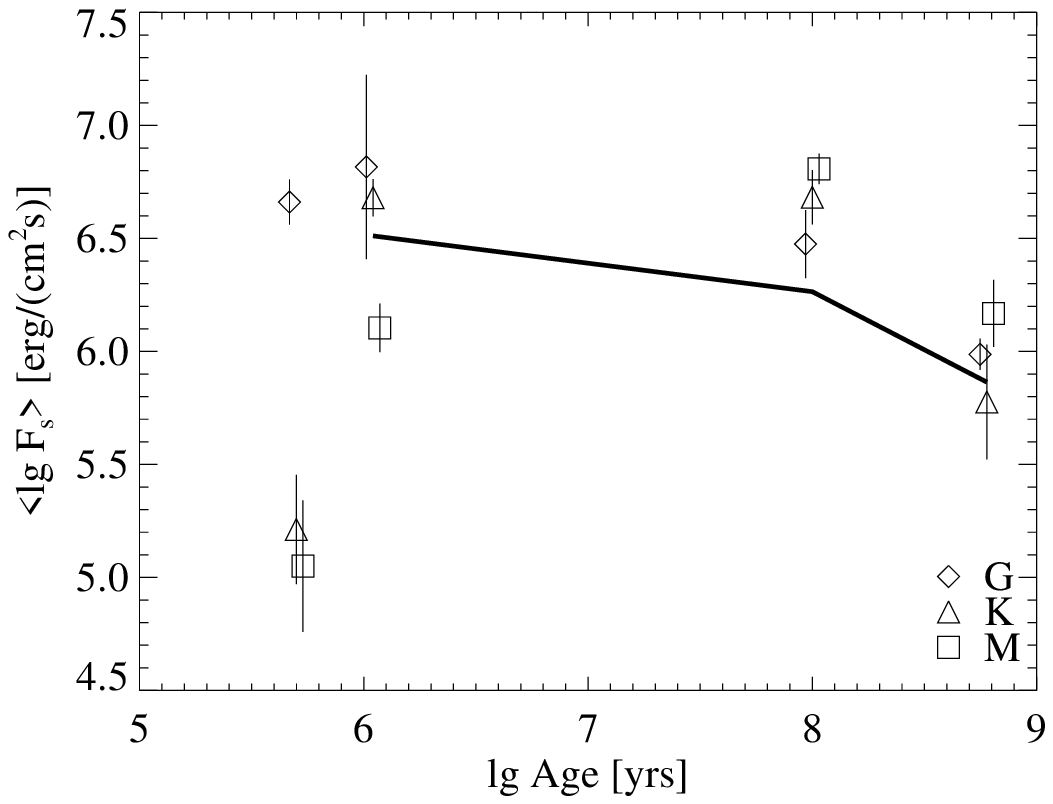}}
\caption{Time evolution of the X-ray surface flux for TTS, Pleiades, and
Hyades for three spectral type groups (plotting symbols for G and M stars
for clarity with a small offset on the age-scale). 
The thick solid line represents a
fit to the mean of $F_{\rm s}$ obtained by combining G, K, and M stars from
the wTTS, Pleiades, and Hyades sample. The
slope of this exponential decay is $-2.0 \pm 0.1$ in agreement with earlier
estimates for smaller samples of stars from the same region (see text).}
\label{fig:Fs_time}
\end{center}
\end{figure}

We have examined the mean level of X-ray surface flux for each 
age group in order to infer a decay law.
In Fig.~\ref{fig:Fs_time} the mean $F_{\rm s}$ is plotted for cTTS, wTTS,
Pleiades, and Hyades, each being split into G, K, and M type stars. 
The X-ray flux increases from cTTS to wTTS as mentioned 
by N95. (Only one cTTS has spectral type of G.)  
An exponential fit to the combined G+K+M sample from the wTTS to the Hyades
age is overlaid, and provides
a slope of $-2.01 \pm 0.09$. This compares well with the result by
\citey{Walter91.1} who found a decrease of $F_{\rm s}$ with $-2.20 \pm
0.21$ for a sample composed of {\em Einstein} detected naked TTS, 
Pleiades and Hyades.

\begin{acknowledgements}
We made use of the Open Cluster Database, compiled by 
C.F. Prosser and J.R. Stauffer. RN wishes to acknowledge financial support 
from the Bundesministerium f\"ur Bildung und Forschung through the
Deutsche Zentrum f\"ur Luft- und Raumfahrt e.V. (DLR)
under grant number 50 OR 0003.
The {\em ROSAT} project is supported by the Max-Planck-Gesellschaft and 
Germany's federal government (BMBF/DLR). We would like to thank the 
referee T. Montmerle for helpful comments and stimulating discussions.
\end{acknowledgements}

\end{document}